\documentclass[preprint,3p,11pt]{elsarticle}

 \usepackage{epsfig}
\usepackage{amssymb}
\usepackage{amsmath}
\def\bma{\left( \begin{array} }
\def\ema{\end{array} \right)}

\title{A light triplet boson and Higgs-to-diphoton
                             in supersymmetric type II seesaw}

\author{Eung Jin Chun}
\ead{ejchun@kias.re.kr}
\author{and Pankaj Sharma}
\ead{pankajs@kias.re.kr}

\address[rvt]{Korea Institute for Advanced Study\\
Heogiro 85, Dongdaemun-gu, Seoul 130-722, Korea}

\begin{document}

\begin{abstract}The supersymmetric type II seesaw may leave a limit where
a triplet boson along with the standard Higgs boson remains light.
Working in this limit with small triplet vacuum expectation vlaues,
we explore how much such a light triplet boson can contribute to the Higgs boson decay
to diphoton, and analyze the feasibility to observe it through
same-sign di-lepton and tetra-lepton signals in the forthcoming LHC run
after setting a LHC7 limit in a simplified parameter space of the triplet vaccum expectation value and the doubly charged boson mass.
\end{abstract}

\maketitle

\section{Introduction}

The origin of neutrino masses and mixing can be attributed to an $SU(2)_L$ triplet boson
which contains a doubly charged boson \cite{type2}.  This scenario of type II seesaw
can be  readily probed at colliders by observing same-sign di-lepton resonances coming
from the doubly charged boson decay.
Furthermore, the observation of the flavor dependent branching ratios of the doubly charged boson
allows us to determine the neutrino mass pattern  \cite{chun03}-\cite{delAguila:2008cj}.
Such signals are being searched for at the LHC and non-observation of them puts a strong
bound on the mass or the di-lepton branching ratio of the doubly charged boson \cite{cmsH++,atlasH++}.

The doubly (and singly) charged boson in type II seesaw can make a sizable contribution
to the Standard Model (SM) Higgs boson decay to diphoton \cite{arhrib11}-\cite{others}. Thus,
the type II seeasaw would be a natural framework for the explanation of the current deviation of the Higgs-to-diphoton rate observed by both ATLAS and CMS \cite{higgs1212}.
Whether or not the current anomaly disappears,
more precise measurement of the diphoton rate in the coming years will place an
additional restriction on the triplet boson mass and coupling to the SM Higgs boson.
As was studied in Ref.~\cite{chun1209}, the scalar couplings of
the triplet boson are tightly constrained by the EWPD, perturbativity and vacuum stability
conditions and thus a sizable deviation of the Higgs-to-diphoton rate can be arranged only by relatively light triplet boson. This implies that the LHC search for a doubly charged boson
in the small mass region remains important as it can hide from the current search
by having a small di-lepton branching fraction.
This is also the case with the supersymmetric type II seesaw model as will be discussed in this work.

In additional to clean same-sign di-lepton signals, the type II seesaw
may exhibit a novel signature of same-sign tetra-leptons which arises from triplet-antitriplet oscillation \cite{chun1206}. A pair of neutral triplet and antitriplet produced from
$pp$ collision can evolve to a pair of two neutral triplets or antitriplets
which then decay to a pair of same-sign doubly charged bosons
leading to four leptons of the same sign in the final states.
 Observation of such events together with same-sign
di-leptons will be a clear confirmation of the existence of not only a doubly charged boson
but also the triplet state of the type II seesaw.

In this paper, we consider a supersymmetric version of the type II seesaw
in a limit where light degrees of freedom consist of one of the triplet bosons
as well as the usual SM Higgs boson as in the non-supersymmetric model.
Although fine-tuned, this would be a unique possibility for a triplet boson
to leave interesting phenomenological impacts in the supersymmetric type II seesaw.
When the triplet Yukawa couplings to leptons in the superpotential are taken to be small,
 which might be indicated by the smallness of neutrino masses,
the mass splitting among the triplet components and their scalar couplings
are determined by the gauge couplings through the $D$-term potential, and thus generically
smaller than in the non-supersymmetric model.
We will show that a sizable contribution
to the Higgs boson decay to diphoton can occur for a very light doubly charged boson.
Although the particle content in this scenario is the same as in the non-supersymmetric
type II seesaw, there appear more parameters relevant for the discussion of
the neutrino mass generation and the triplet decays. Taking this difference into account,
we will analyze the LHC8 and LHC14 reach of same-sign di and tetra lepton signals in the parameter
space of the doubly charged boson mass and the ratio between the triplet and doublet vacuum
expectation values.

\section{Triplet boson spectrum and couplings}

The supersymmetric type II seesaw model introduces a vector-like pair of $SU(2)_L$ triplets: $\Delta=(\Delta^{++}, \Delta^{+}, \Delta^{0})$
and $\bar{\Delta}=(\bar{\Delta}^{0}, \bar{\Delta}^{-}, \bar{\Delta}^{--})$ with $Y=1$
and $Y=-1$, respectively. In the matrix representation, they are written as
\begin{equation}
 {\bf \Delta} = \begin{pmatrix} {\Delta^+\over\sqrt{2}} & \Delta^{++} \cr
                          \Delta^0 & -{\Delta^+\over\sqrt{2}}
                \end{pmatrix} \,, \quad
 \bar{\bf \Delta} = \begin{pmatrix} {\bar{\Delta}^-\over\sqrt{2}} & \bar{\Delta}^{0} \cr
                          \bar{\Delta}^{--} & -{\bar{\Delta}^-\over\sqrt{2}}
                          \end{pmatrix} \,.
\end{equation}
Then, the gauge-invariant superpotential contains
\begin{equation}
 W = {1\over2} f_{ij} L_i^T i\tau_2 {\bf \Delta} L_j + {1\over2} \lambda_1 H_1^T i\tau_2
 \mathbf{\Delta} H_1
 - {1\over2} \lambda_2
 H_2^T i\tau_2 \bar{\mathbf{\Delta}} H_2 + \mu H_1^T i\tau_2 H_2 + M {\rm Tr}[{\bf \Delta}\bar{\bf \Delta}] \,.
\end{equation}
Notice that the minimization of the scalar potential gives rise to non-trivial vacuum expectation values of the triplets generating  the neutrino mass matrix
\begin{equation}
 M^\nu_{ij} = f_{ij} \langle \Delta^0 \rangle.
\end{equation}
Non-vanishing triplet vacuum expectation values arise from the couplings $\lambda_{1,2}$, and thus
the neutrino masses around 0.1 eV requires $f\lambda_{1,2} \sim 10^{-12}$. In this work, we will assume the smallness of both couplings: $f, \lambda_{1,2} \ll 1$.
The scalar potential relevant for our discussion is presented in Appendix.

Let us first calculate the Higgs triplet boson spectrum before considering
the mixing mass between the Higgs doublet and triplet bosons.
Ignoring the contribution of the triplet vacuum expectation values,
$ \langle \Delta^0 \rangle \equiv v_{\Delta}/\sqrt{2}$
and $\langle \bar\Delta^0 \rangle \equiv  v_{\bar\Delta}/\sqrt{2}$,
we get the  mass matrix of the component $\Delta^a$ and $\bar{\Delta}^{\bar a}$:
\begin{equation} \label{Fmass}
 \begin{pmatrix} M^2_{\Delta^a} & B_M \cr
 B_M & M^2_{\bar{\Delta}^{\bar{a}}} \end{pmatrix}
\end{equation}
where $B_M$ is a dimension-two soft mass, and
$a$ labels $++$, $+$ or $0$.  Here the diagonal components are given by
\begin{eqnarray}
 M^2_{\Delta^{a}}
 &\equiv& M^2 + m_{\Delta}^2 + d_a M_Z^2 c_{2\beta} + {c_a\over4} \lambda_1^2 v_0^2 c_\beta^2,
 \\
 M^2_{\bar\Delta^{\bar a}}
 &\equiv& M^2 + m_{\bar\Delta}^2 - d_a   M_Z^2 c_{2\beta} + {c_a\over4} \lambda_2^2 v_0^2 s_\beta^2
 \nonumber
\end{eqnarray}
where $(d_{++}, d_+, d_0) = (1-2s_W^2, -s_W^2, -1)$,  $(c_{++}, c_+, c_0 ) = (0,1,2)$, and
$\langle H^0_1 \rangle \equiv v_0 c_\beta/\sqrt{2}$ and $\langle H_2^0 \rangle \equiv v_0 s_\beta/\sqrt{2}$.
The triplet mass matrix can be diagonalized by the rotation:
\begin{eqnarray} \label{Udiag}
\Delta^{a} &=& c_{\delta^a} \Delta^{a}_1 - s_{\delta^a} {\Delta}^{a}_2
 \\
\bar\Delta^{a *} &=& s_{\delta^a} \Delta^{a}_1 + c_{\delta^a} {\Delta}^{a}_2
\nonumber
\end{eqnarray}
where the rotation angle $\delta^a$ satisfies the relation $t_{2\delta^a} = 2 B_M/ (M^2_{\Delta^a}-M^2_{\bar\Delta^{\bar a}})$.

Now taking the limit of vanishingly small $\lambda_{1,2}$ ensuring
 tiny triplet vacuum expectation values, we get the mass eigenvalues
given by
\begin{equation}
 M_{\Delta^a_{1,2}}^2 = {1\over2} \left[
 2 M^2 + m^2_{\Delta} + m^2_{\bar\Delta} \mp
 \sqrt{ (m^2_\Delta - m^2_{\bar\Delta} + 2d_a M_Z^2 c_{2\beta})^2 + 4 B_M^2} \right],
\end{equation}
and thus the mass splitting among the triplet components is controlled by the $D$ term.
In the leading order of the $D$ term contribution (assuming $|m^2_\Delta - m^2_{\bar\Delta}| \gg M_Z^2$), we get the relation
\begin{eqnarray} \label{Msplit}
M^2_{\Delta^{0}_{1,2}} - M^2_{\Delta^{+}_{1,2}}  = M^2_{\Delta^{+}_{1,2}} - M^2_{\Delta^{++}_{1,2}}
 = \mp (1-s_W^2) c_{2\delta}c_{2\beta}  M_Z^2
\end{eqnarray}
where
\begin{equation} \label{deltais}
 c_{2\delta} \equiv
 - {m^2_\Delta - m^2_{\bar\Delta} \over
    \sqrt{ (m_\Delta^2 - m^2_{\bar\Delta})^2 + 4 B_M^2} } .
\end{equation}
In the same limit, the three rotation angles $\delta^a$ can be approximated by the angle
$\delta$ (\ref{deltais}): $\delta^a \approx \delta$.
Thus, one can take $M_{\Delta_{1,2}^{++}}$ and $\delta \approx \delta^a$ as input parameters
to determine the other masses $M_{\Delta^{+,0}_{1,2}}$
through the simple relation (\ref{Msplit}).
This leads to the mass hierarchy
\begin{equation}
 M_{\Delta^{++}_1} < M_{\Delta^{+}_1} < M_{\Delta^{0}_1} <
 M_{\Delta^{0}_2} < M_{\Delta^{+}_2} < M_{\Delta^{++}_2}
\end{equation}
for
$c_{2\delta}c_{2\beta}<0$, which has a similar pattern as in
the non-supersymmetric type II seessaw \cite{chun03}.
  Here let us note that the lighter
triplet state can be made much lighter than the heavier state if there is a sizable
cancellation between the positive and negative contributions in the mass-squared eigenvalue
$M^2_{\Delta_1}$. Such a fine-tuned limit (i.e., $M_{\Delta_1} \ll M_{\Delta_2}$)
is assumed in the major part of this work.

From the minimization conditions of the scalar potential (\ref{VFDs}), one can calculate
$ \xi_\Delta \equiv v_\Delta / v_0$ and $ \xi_{\bar\Delta} \equiv v_{\bar{\Delta}} / v_0$.
Following the diagonalization matrix (\ref{Udiag}), it is convenient to split the ratio $\xi_\Delta$ and $\xi_{\bar\Delta}$ into $\xi_1$ and $\xi_2$:
\begin{eqnarray}
 \xi_\Delta &=&
 c_{\delta^0} \xi_{1} - s_{\delta^0} \xi_2,
 \\
 \xi_{\bar\Delta} &=&
 s_{\delta^0} \xi_1 + c_{\delta^0} \xi_2 , \nonumber
\end{eqnarray}
where $\xi_{1,2}$ are given by
\begin{eqnarray}
\xi_1 &\equiv&
 { v [ (\lambda_1 s_{\delta^0} c_\beta^2 + \lambda_2 c_{\delta^0} s_\beta^2)M
+  (\lambda_1 c_{\delta^0} + \lambda_2 s_{\delta^0}) \mu s_{2\beta}
+ \lambda_1 A_1 c_{\delta^0} c_\beta^2 + \lambda_2 A_2 s_{\delta^0} s_\beta^2 ]
\over 2\sqrt{2} M^2_{\Delta^0_1} } ,
 \nonumber\\
\xi_2 &\equiv&
{ v [ (\lambda_1 c_{\delta^0} c_\beta^2 - \lambda_2 s_{\delta^0} s_\beta^2) M
-  (\lambda_1 s_{\delta^0} - \lambda_2 c_{\delta^0}) \mu s_{2\beta}
- \lambda_1 A_1 s_{\delta^0} c_\beta^2 + \lambda_2 A_2 c_{\delta^0} s_\beta^2]
 \over 2 \sqrt{2} M_{\Delta^0_2}^2} .
\nonumber
\end{eqnarray}
Recall that
the $\rho$ parameter constraint,
$\rho-1 \lesssim 0.1 \%$, puts a rough bound $|\xi| \lesssim 1\%$. We will further assume $|\xi| \ll 0.01$ for which the mixing between the Higgs doublet and triplet can be safely ignored and thus the
the triplet states $\Delta^a_{1,2}$ can be taken as the full mass eigenstates to a good approximation.

\medskip

The doubly charged bosons have the Yukawa couplings to di-lepton and
the gauge couplings to di-$W$ given by
\begin{eqnarray} \label{ll-WW}
{\cal L} &=& {1\over\sqrt{2}} \left[ c_\delta f_{ij} \bar l^c_i P_L l_j
+ g \xi_1 M_W W^- W^-\right] \Delta^{++}_1 + h.c. \\
&+&  {1\over\sqrt{2}} \left[ -s_\delta f_{ij} \bar l^c_i P_L l_j
+ g \xi_2 M_W W^- W^-\right] \Delta^{++}_2
+ h.c.. \nonumber
\end{eqnarray}
There are also the scalar couplings to the (heavy) charged Higgs boson $H^\pm$  [see Appendix]
which will be ignored in our analysis.
Thus, the light doubly charged boson $\Delta_1^{++}$ (lighter than $H^\pm$) can decay
only to $l^+ l^+$ and $W^+ W^+$ depending on the corresponding couplings $c_\delta f_{ij}$ and $\xi_1$.

Throughout this work, we will take the decoupling limit of the heavy pseudo-scalar and
charged bosons from the Higgs doublet and thus use the relation:
$H_1^0 =c_\beta (v_0 + h )/\sqrt{2} $, and $H_2^0 =s_\beta (v_0 + h )/\sqrt{2} $.
Another important aspect of the supersymmetric type II seesaw is that
the doubly (singly) charged boson couplings to the Standard Model Higgs boson $h$
arises from the $D$-term potential:
\begin{equation} \label{hDD}
V_D =c_{2\beta} c_{2\delta} v_0\, h
\left[ {g^2 -g^{\prime 2} \over 2} (|\Delta^{++}_1|^2 -|\Delta^{++}_2|^2)
 -{g^{\prime 2}\over2} (|\Delta^{+}_1|^2 -|\Delta^{+}_2|^2) \right].
\end{equation}
The effect of these couplings to one-loop diagrams for the Higgs boson decay to di-photon
will be discussed in the next section.

\section{Higgs-to-diphoton rate}

\begin{figure}
\begin{center}
\includegraphics[width=3.2in]{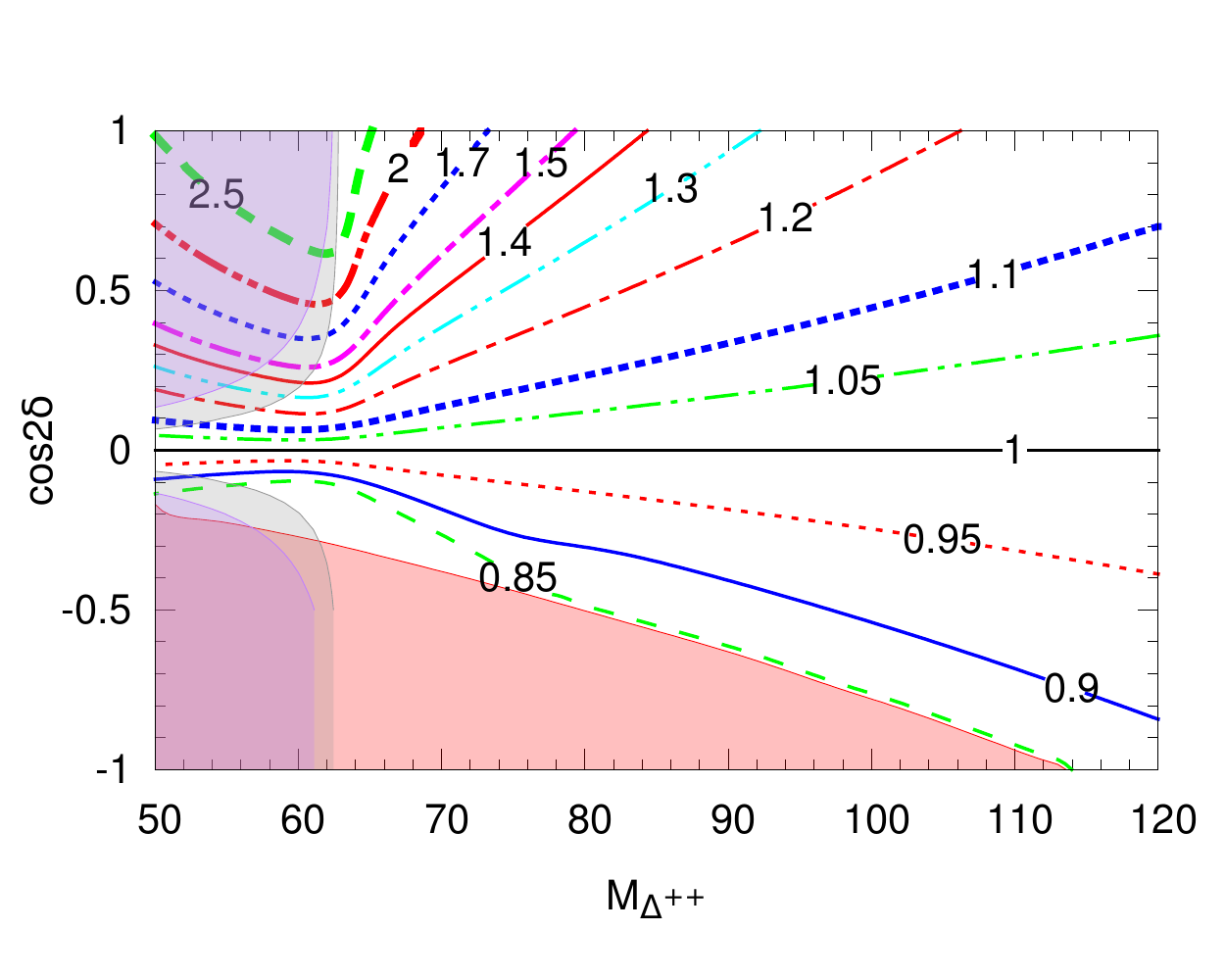}
\caption{The $R_{\gamma\gamma}$ contours in the ($M_{\Delta^{++}}$,$c_{2\delta}$) plane for $t_\beta=10$. The pink region in the below is
disallowed by positivity of the triplet masses. 
 In the purple and gray regions on the left, 
 BF($h\to \Delta^{++}\Delta^{--}$) becomes larger than 30 \% and 10 \%,  respectively.} \label{Rgg}
\label{polmh}
\end{center}
\end{figure}

Including the contribution from the couplings (\ref{hDD}),
one obtains the following decay rate of the Higgs boson to diphoton \cite{djouadi05}:
\begin{eqnarray} \label{hgg}
\Gamma(h\to \gamma\gamma) &=& {G_F \alpha^2 m_h^3\over 128\sqrt{2}\pi^3}
\left| \sum_f N_c Q_f^2\, g^h_{ff} A^h_{1/2}(x_f) + g^h_{WW} A^h_1(x_W) \right.\\
&& \left. +  g^h_{\Delta^+ \Delta^-} [  B^h_0(x_{\Delta_1^+})
- B^h_0(x_{\Delta_2^+})]
+ 4g^h_{\Delta^{++} \Delta^{--}} [B^h_0(x_{\Delta_1^{++}})
- B^h_0(x_{\Delta_2^{++}})]\right|^2 \nonumber
\end{eqnarray}
where $x_i = m_h^2/4 m_i^2$. The loop functions are defined by
\begin{eqnarray}
A^h_{1/2}(x) &=& 2x^{-2}[x + (x-1) f(x)] \\
A^h_{1}(x) &=& -x^{-2}[2x^2 + 3 x + 3 (2x-1) f(x)] \nonumber\\
B^h_0(x) &=& -4 x^{-1}[x-f(x)] \nonumber\\
\quad\mbox{where}&&\!\!\!\! f(x) =
 \begin{cases}
 \arcsin^2\sqrt{x} \quad\mbox{for}\quad x\leq 1 \cr
 - {1\over 4} \left[ \ln{ 1+\sqrt{1-x^{-1}} \over 1-\sqrt{1-x^{-1}}} - i\pi\right]^2
\quad\mbox{for}\quad x>1 \cr \end{cases} \nonumber
\end{eqnarray}
The couplings appearing in (\ref{hgg}) are as follows:
$g^h_{ff} =1$ for the top, $g^h_{WW}=1$ for the $W$ boson, and
\begin{equation}
 g^h_{\Delta^+ \Delta^+} = - {c_{2\delta} c_{2\beta}t_W^{2} } {  M_W^2 \over m_h^2} \quad\mbox{and}\quad
 g^h_{\Delta^{++} \Delta^{++}} = {c_{2\delta} c_{2\beta} (1-t_W^{2}) } { M_W^2 \over m_h^2},
\end{equation}
for the singly and doubly charged triplet bosons, respectively.
Notice that these couplings are determined by the $D$-term potential and generically smaller than those coming from the scalar potential in the non-supersymmetric type II seesaw model
constrained by EWPD, perturvativity and vacuum stability \cite{chun1209}.
Furthermore, the light and heavy triplet bosons give opposite contributions.
Thus, a sizable deviation to the SM prediction on the diphoton rate can be obtained if the heavy triplet boson decouples away and the light triplet boson becomes even lighter than the Higgs boson.
 Furthermore,
 the contributions from $\Delta_1^{++}$ and $\Delta_1^{+}$ give a constructive interference with the SM contribution (which is about $-6.5$)
 for $c_{2\delta} c_{2\beta} <0 $ corresponding to the mass hierarchy  $M_{\Delta_1}^{++} < M_{\Delta_1}^{+} < M_{\Delta_1}^{0}$.
 For these reasons we will consider the cases with $M_{\Delta_1} \ll M_{\Delta_2}$, and $c_{2\delta}>0$ (with $c_{2\beta}<0$)
 to maximize the triplet boson contribution to the diphoton rate.
Recall that, given $c_{2\delta}$ and $M_{\Delta_1}^{++}$, $M_{\Delta^+_1}$ is determined by
$
 M^2_{\Delta_1^+} \approx  M_{\Delta_1^{++}}^2 - c_{2\beta} c_{2\delta} (1-s_W^2) M_Z^2$
  (\ref{Msplit}).

In Fig.~\ref{Rgg}, we plot the deviation of the diphoton rate from the SM value in the plane of
$M_{\Delta^{++}}$ and $c_{2\delta}$. Here, $R_{\gamma\gamma}$ denotes the ratio between the Higgs-to-diphoton rates in the type II seesaw and in the SM:
$R_{\gamma\gamma} \equiv \Gamma(h \to \gamma\gamma)_{\rm II}/\Gamma(h\to \gamma\gamma)_{\rm SM}$.
Hereafter $\Delta_1$ will be denoted by $\Delta$ dropping the subscript.  As can be seen, one can 
have a sizable enhancement of the diphoton rate only in a limited region of small $M_{\Delta^{++}}$ and large $c_{2\delta}$.  
In this region, the Higgs coupling to the triplet boson becomes large enough to make
 the Higgs decay $h\to \Delta^{++}\Delta^{--}$ comparable to, e.g., the standard decay of 
 $h\to WW$ as both of them
come from the gauge vertices. To see this effect, Fig.~\ref{Rgg} also shows
the contour lines for which the
branching fraction (BF) of the $h\to \Delta^{++}\Delta^{--}$ decay becomes 30 \% and 10 \%.
Although consistent with the SM prediction, the current data \cite{higgs1212} is not precise enough
to rule out such non-standard Higgs decays. Therefore, it would be interesting to contemplate observing the 
doubly charged boson (and its companions) from the Higgs decay.  
The final states from the Higgs decay consist of softer leptons and 
jets and thus would be distinguishable from the conventional ones.
The precise significance of observing such an exotic Higgs properly
needs a detailed simulation which we leave for a future work.

\section{Same-sign di-/tetra-lepton signatures}

The doubly charged boson in the type II seesaw
is directly searched for by looking at same-sign di-lepton resonances
from the decay $\Delta^{\pm\pm} \to l^\pm l^\pm$. No excess over the background expectation
has been observed so far and limits are placed on the doubly charged boson mass
depending on the branching ratio assumed for the di-lepton channel \cite{cmsH++,atlasH++}.

\begin{figure}
\begin{center}
\includegraphics[width=3.in]{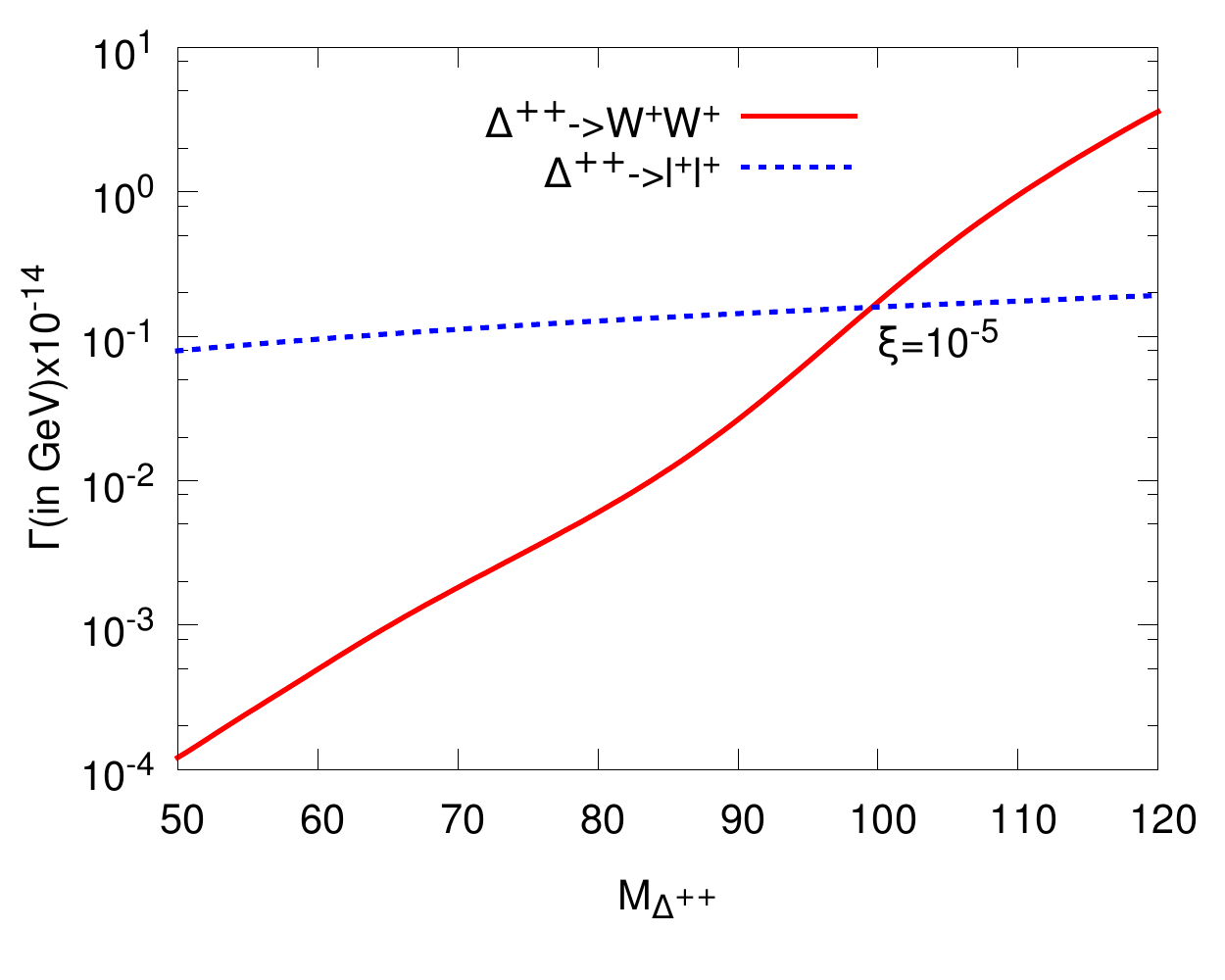} 
\caption{The decay rates $\Gamma_{WW}$ and $\Gamma_{ll}$
of  $\Delta^{++}$ for $\xi\equiv \xi_1 = \xi_\Delta/c_\delta=10^{-5}$.}
\label{hppWW}
\end{center}
\end{figure}

When the doubly charged boson is lighter than the singly charged and neutral components ($c_{2\delta}>0$), it decays to either di-leptons or di-$W$
through the coupling in (\ref{ll-WW}).
The di-lepton decay rates are then given by
\begin{equation}
\Gamma_{l_i l_j} \equiv  \Gamma(\Delta^{++} \to l^+_i l^+_j) = S c_{\delta}^2 {|f_{ij}|^2 \over 16\pi} M_{\Delta^{++}}
 \end{equation}
where $S=2 (1)$ for $i\neq j (i=j)$. From the neutrino mass relation:
$M^\nu_{ij} = f_{ij} \xi_\Delta v_0$, one gets the total di-lepton rate \cite{chun03}:
\begin{equation}
\Gamma_{ll} \equiv \sum_{i,j} \Gamma_{l_i l_j} =  {1\over 16\pi} {c_\delta^2{\bar m_\nu}^2 \over |\xi_\Delta|^2 v_0^2} M_{\Delta^{++}}
\end{equation}
where $\bar m_\nu^2 = \sum_i m_{\nu_i}^2$ is the sum of three neutrino mass-squared eigenvalues.
On the other hand, the di-$W$ decay $\Delta^{++} \to W^{+} W^{+}$, where one or both of $W$'s are off-shell for the range of $M_{\Delta^{++}}$ considered in this work, depends on
the parameter $\xi_1$, that is, $\Gamma_{WW} \propto |\xi_1|^2$. In Fig.~\ref{hppWW}, the decay rates, $\Gamma_{WW}$ and $\Gamma_{ll}$, of the doubly charged boson are
presented taking $\xi\equiv \xi_\Delta/c_\delta=\xi_1=10^{-5}$ for comparison.

\begin{figure}[h]
\begin{center}
\hspace{-0.8cm}
\includegraphics[width=3.in]{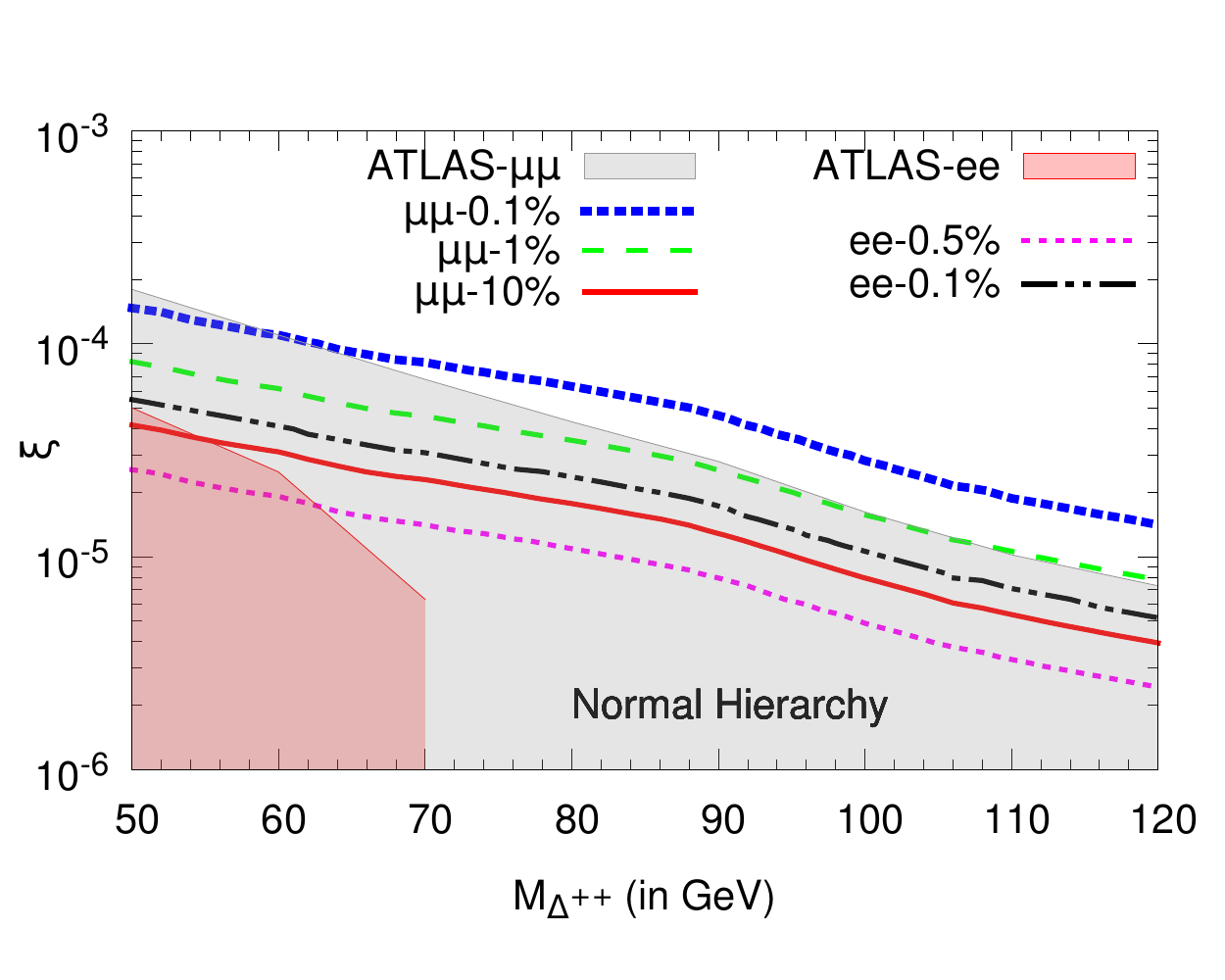} \hspace{-0.42cm}
\includegraphics[width=3.in]{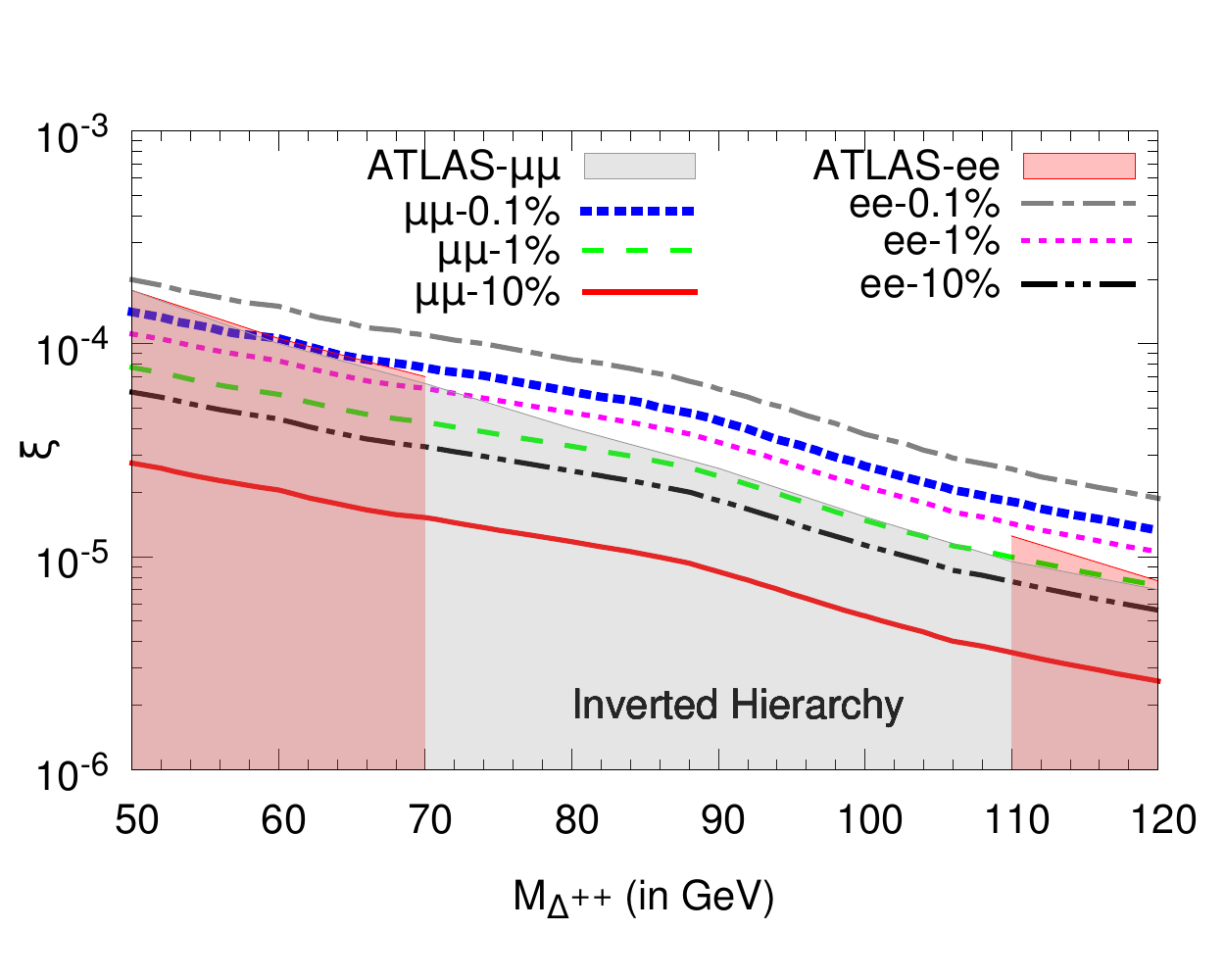} 
\caption{BF($\mu\mu$) and BF($ee$) in the ($M_{\Delta^{++}}$, $\xi$) plane for NH (left) and IH-hierarchy (right),
and the ATLAS exclusion regions for $ee$- and $\mu\mu$-channels.}
\label{brhpp}
\end{center}
\end{figure}

\begin{figure}[h]
\begin{center}\hspace{-0.25cm}
\includegraphics[width=2.in]{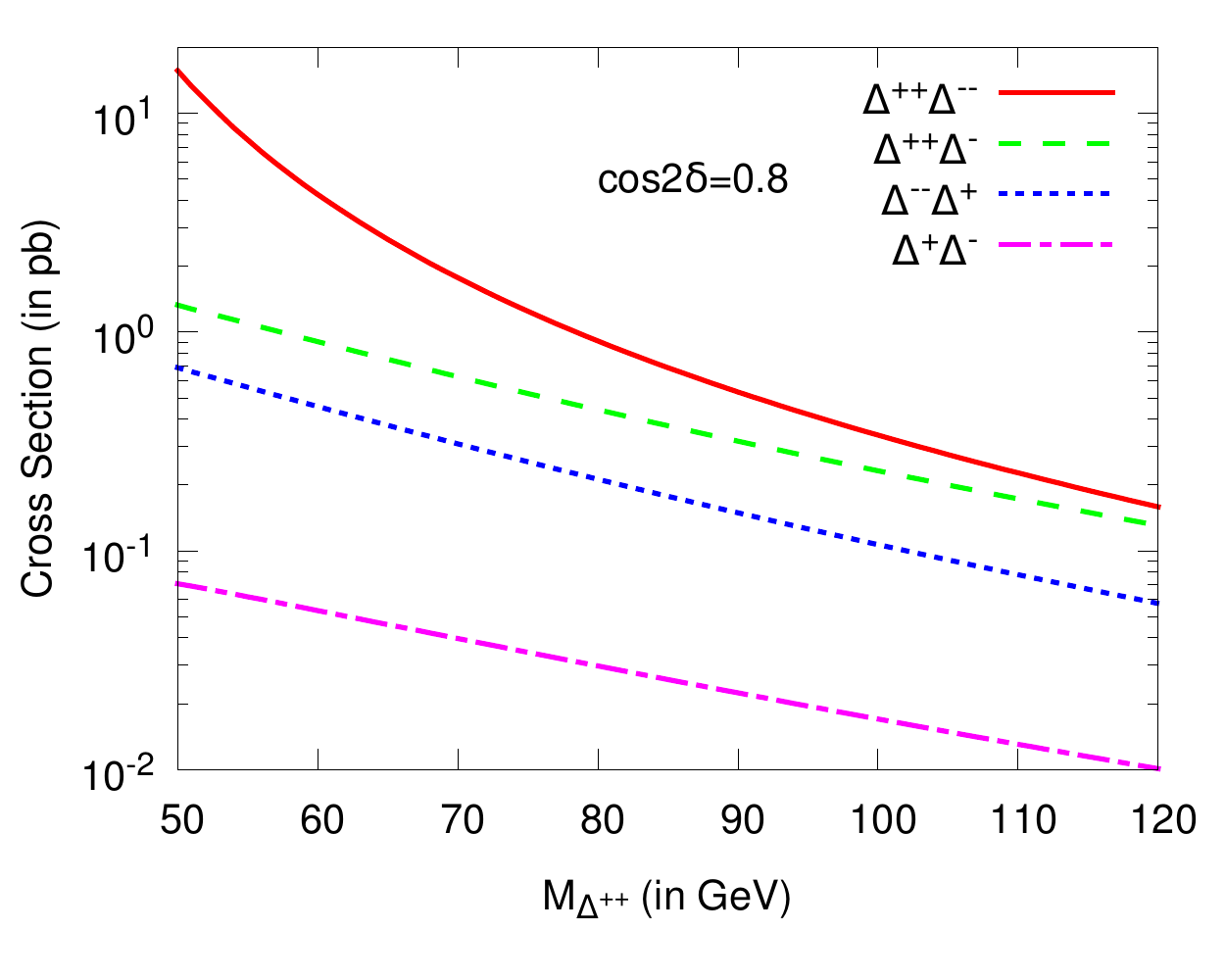}\hspace{-0.2cm}
\includegraphics[width=2.in]{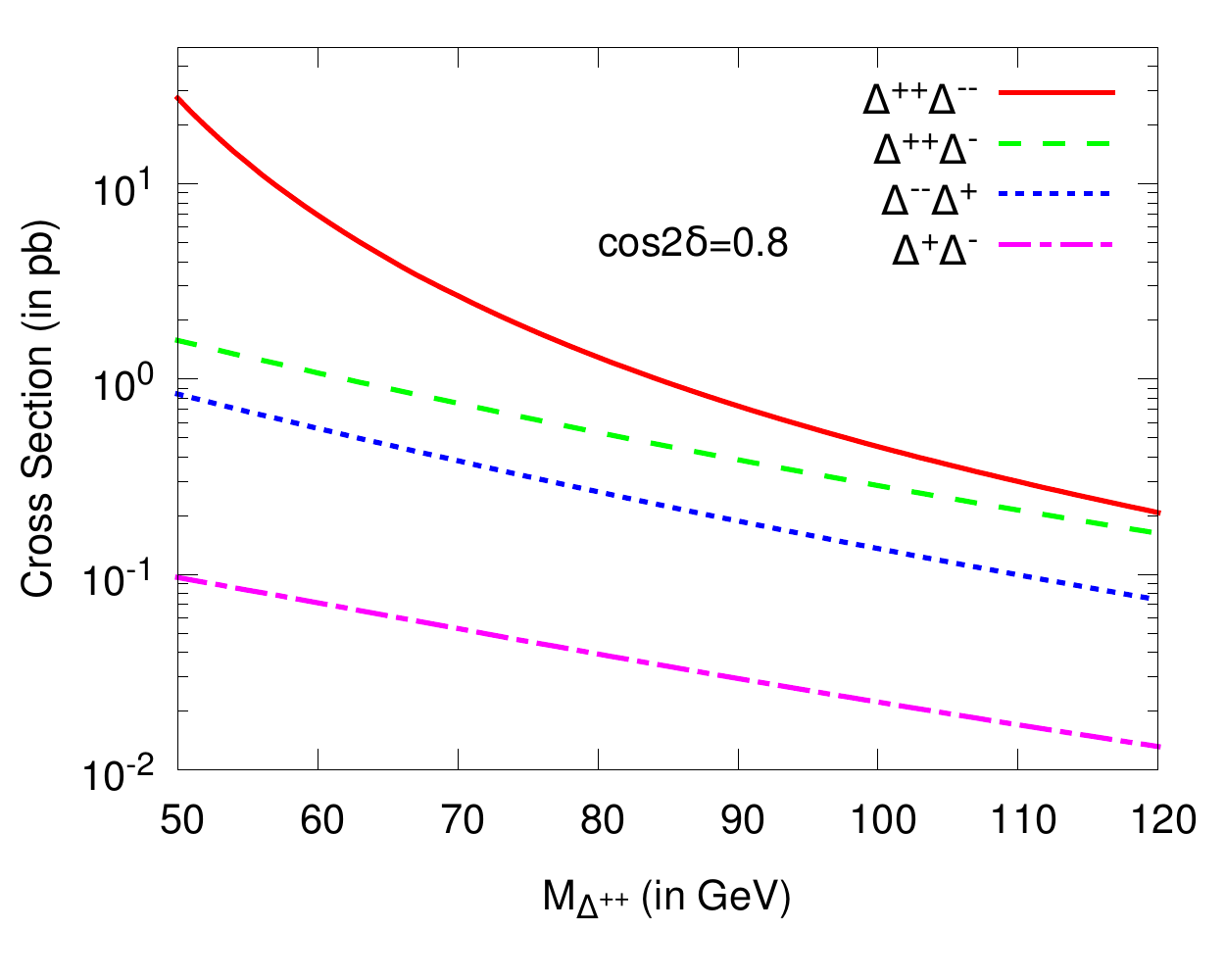}\hspace{-0.2cm}
\includegraphics[width=2.in]{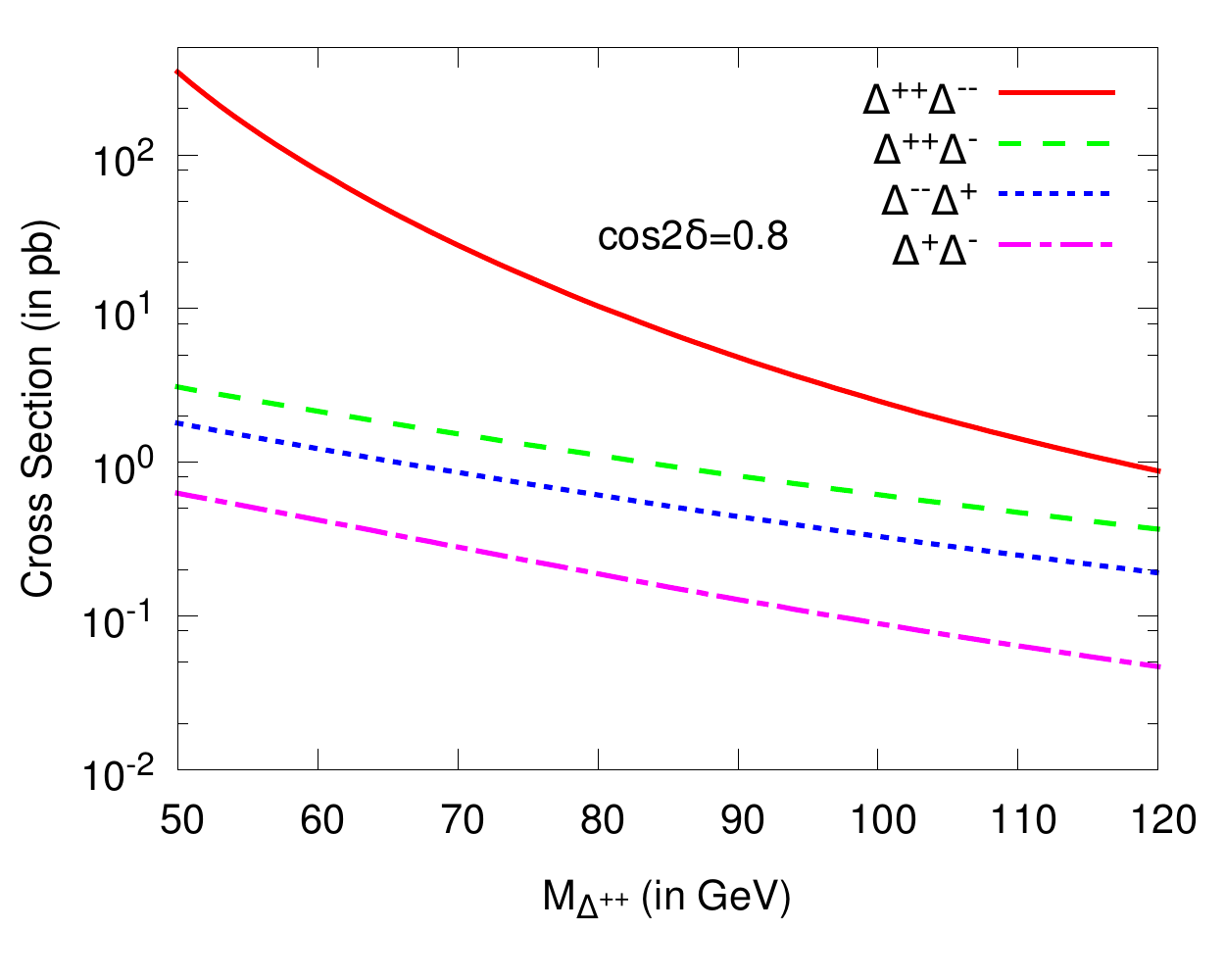}
\includegraphics[width=2.in]{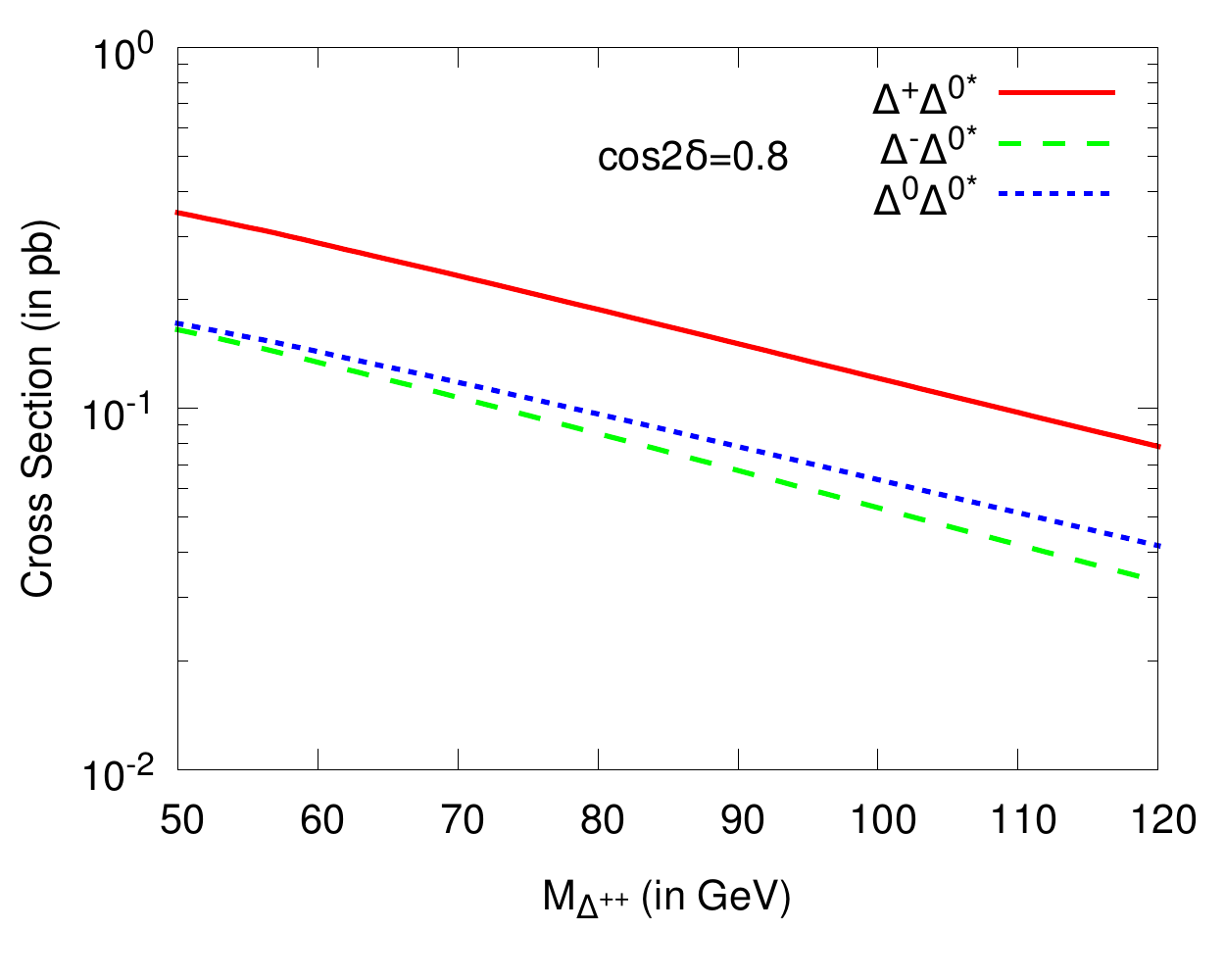}\hspace{-0.2cm}
\includegraphics[width=2.in]{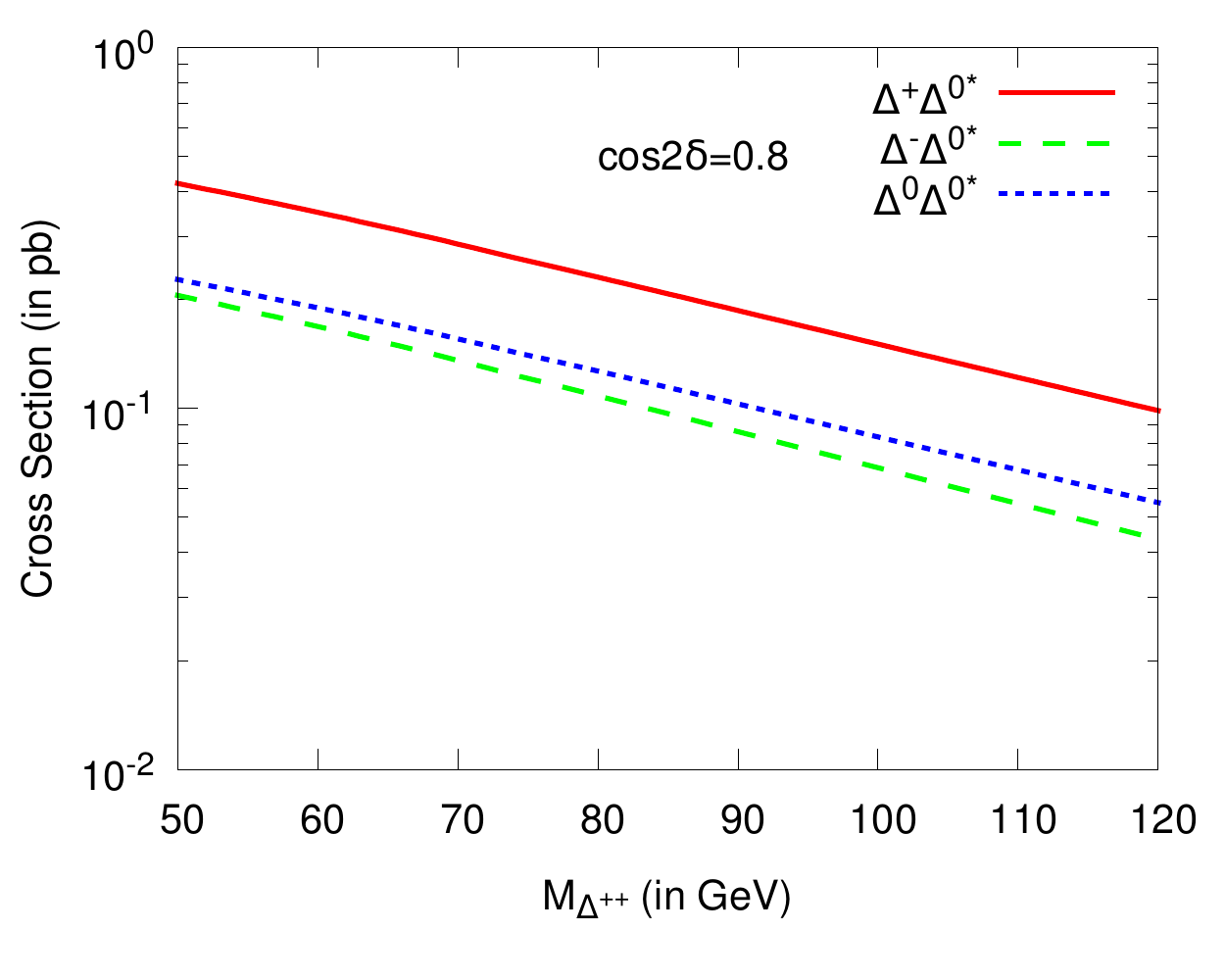}\hspace{-0.2cm}
\includegraphics[width=2.in]{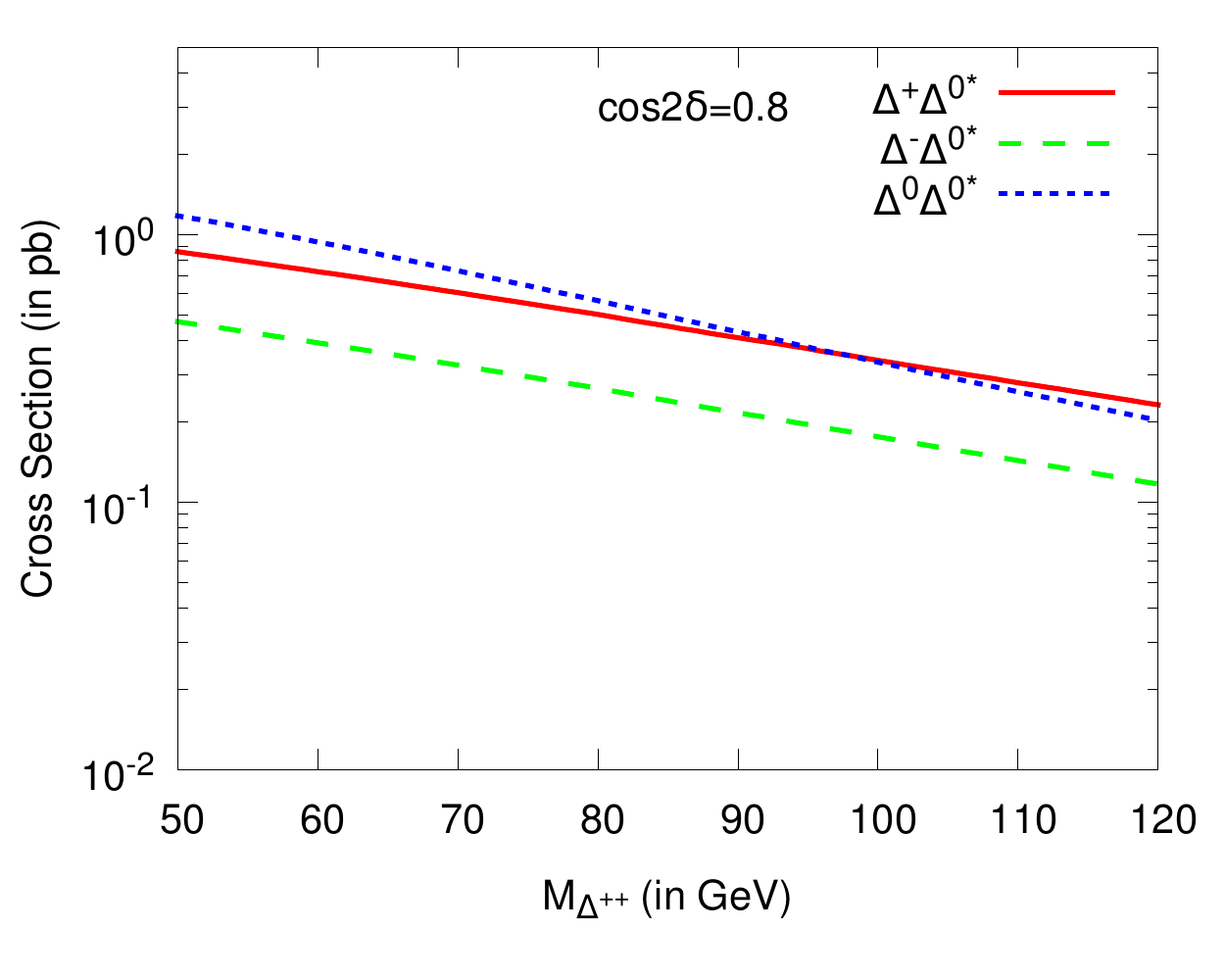}
\caption{All the pair and associated production cross-sections of triplets at 7 TeV (left), 8 TeV (middle) and 14 TeV (right) for $c_{2\delta}=$ 0.8.}
\label{crx-ne} \label{crx-8}
\end{center}
\end{figure}

The current neutrino oscillation data allow us to determine the neutrino mass matrix up to CP phases
and mass hierarchies. From the neutrino mass matrices for the normal (NH) and
inverted (IH) hierarchies \cite{chun1206}, one can find the individual di-lepton decay rate
$\Gamma_{l_i l_j}$  normalized by the total leptonic decay rate $\Gamma_{ll}$ as
\begin{equation} \label{BRij}
\begin{tabular}{|c|c|c|c|c|c|c|}
\hline
$\Gamma_{l_i l_j}/\Gamma_{ll}$ (\%) &  $ee$ & $e\mu$  & $e\tau$ & $\mu\mu$  & $\mu\tau$ & $\tau\tau$ \\
\hline
NH & 0.62 & 5.11 & 0.51 & 26.8  & 35.6 & 31.4 \\
\hline
IH
 & 47.1 & 1.27 & 1.35 & 11.7  & 23.7 & 14.9 \\
\hline
\end{tabular}
\end{equation}
The current LHC search limits on the doubly charged boson \cite{cmsH++,atlasH++}
imply that the total leptonic decay rate $\Gamma_{ll}$ is much smaller than
$\Gamma_{WW}$ for the low mass region.
For $M_{\Delta^{++}} = 70$ GeV, for instance,
the ATLAS limits on $\sigma(pp\to \Delta^{++}\Delta^{--}) \times \mbox{BF}(ee, \mu\mu)$
 \cite{atlasH++} put the upper bounds on the branching fractions;
BF($ee$) $<$ 0.5 \% and BF($\mu\mu$) $<$ 0.2 \% given the production cross-section
$\sigma(pp\to \Delta^{++}\Delta^{--}) \approx 2$ pb at LHC7 [see Fig.~\ref{crx-8}].
Here the individual leptonic branching ratio is given by
$\mbox{BF}(l_i l_j) = \Gamma_{l_i l_j}/(\Gamma_{ll}+\Gamma_{WW})$. Such limits on the di-leptonic
branching fractions translate into $\Gamma_{ll}/\Gamma_{WW} \lesssim 0.01$ as can be seen
in (\ref{BRij}).
From Fig.~\ref{hppWW}, one gets the relation
\begin{equation} \label{ll/WW}
{\Gamma_{ll} \over \Gamma_{WW} } \approx 0.012
\left( 8\times10^{-5} \over \xi_\Delta/c_\delta\right)^2
\left( 8\times10^{-5} \over \xi_1\right)^2
\end{equation}
for  $M_{\Delta^{++}} = 70$ GeV, and  thus the ATLAS search excludes the region, e.g., $\xi_\Delta/c_\delta, \xi_1 \lesssim 8\times 10^{-5}$.
A general analysis in the parameter plane of ($M_{\Delta^{++}}$, $\xi$) with $\xi= \xi_1=\xi_\Delta/c_\delta$,  is made in
Fig.~\ref{brhpp} which shows contours of branching fractions for the decays $\Delta^{\pm\pm}\to e^\pm e^\pm/\mu^\pm \mu^\pm$  in the case of  the normal (left) and inverted (right) hierarchies.
The shaded regions are excluded by ATLAS search for the same-sign di-leptons
$ee$ and $\mu\mu$ coming only from the pair production of $pp \to \Delta^{++}\Delta^{--}$.
The exclusion lines are obtained by smoothing out the fluctuating mass dependence
of the observed limits.
Recall that the doubly charged boson search through the $ee$ channel is excluded for
the mass range 70--110 GeV in the ATLAS analysis due to a large background  coming
from $Z\to e^+ e^-$.
As is clear from (\ref{BRij}), the exclusion is dominated by the $\mu\mu$ channel in the NH case,
whereas the $ee$ channel wherever applicable provides a little more stringent limit in the IH case.
One can see that  BF$(\mu\mu)\gtrsim $0.1\% is excluded  in the low mass region ($M_{\Delta^{++}}<70$ GeV) and the limit gets more relaxed up to BF($\mu\mu) \sim 1$ \% for $M_{\Delta^{++}}\sim 100$ GeV,
which correspond to the lower limits; $\xi \gtrsim 10^{-4}$ and $10^{-5}$, respectively.
There is still a lot of parameter space available for such a light doubly charged boson
to escape the current LHC search and waiting for further searches.

\begin{figure}[h]
\begin{center}
\includegraphics[width=3.0in]{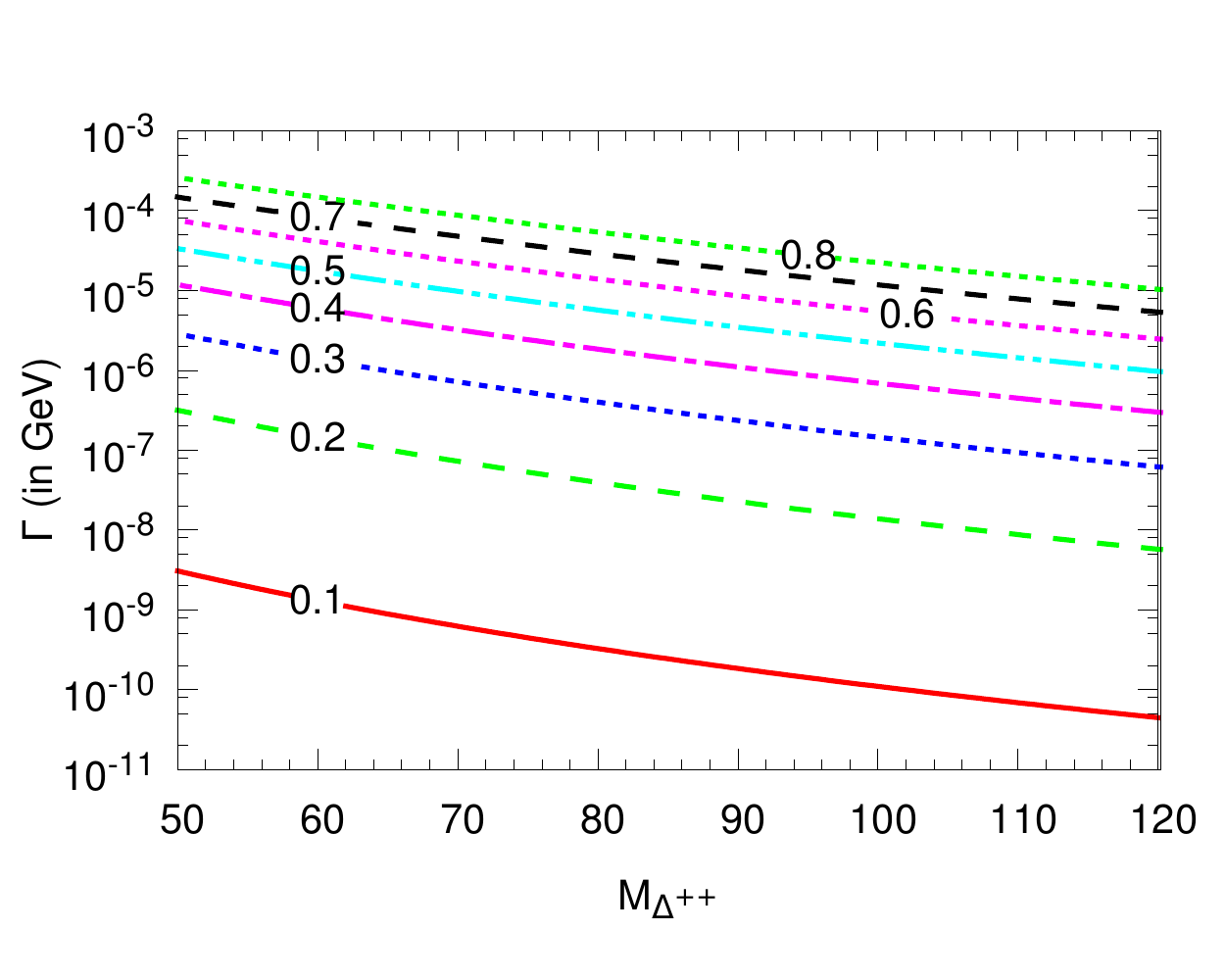}
\caption{The decay rate $\Gamma(\Delta^0\to \Delta^+ W^{-*})$ as a function of $M_{\Delta^{++}}$ for different values of $c_{2\delta}=0.1 - 0.8$.}
\label{gamh0}
\end{center}
\end{figure}

\begin{figure}[h]
\begin{center}\hspace{-0.25cm}
\includegraphics[width=2.in]{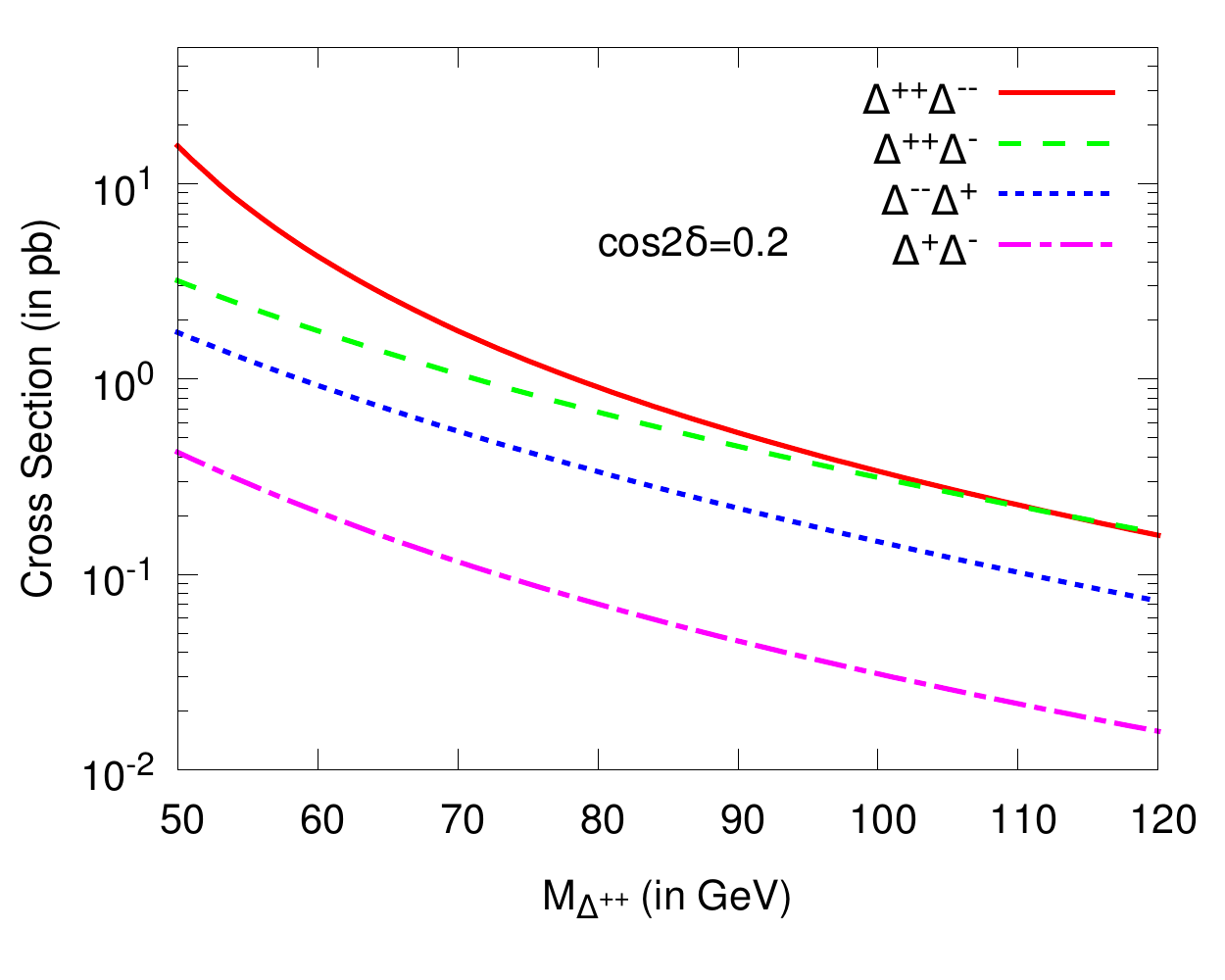}\hspace{-0.2cm}
\includegraphics[width=2.in]{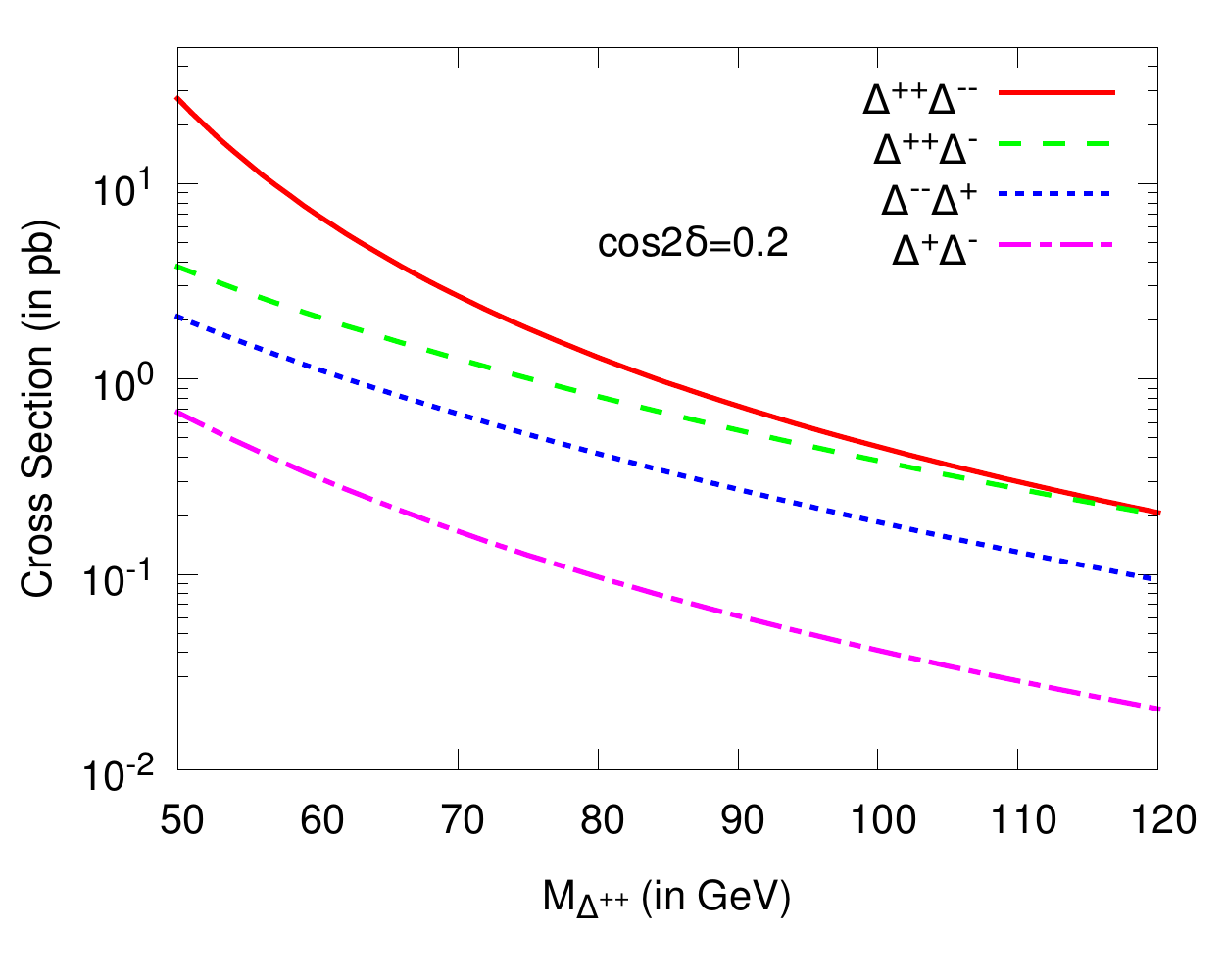}\hspace{-0.2cm}
\includegraphics[width=2.in]{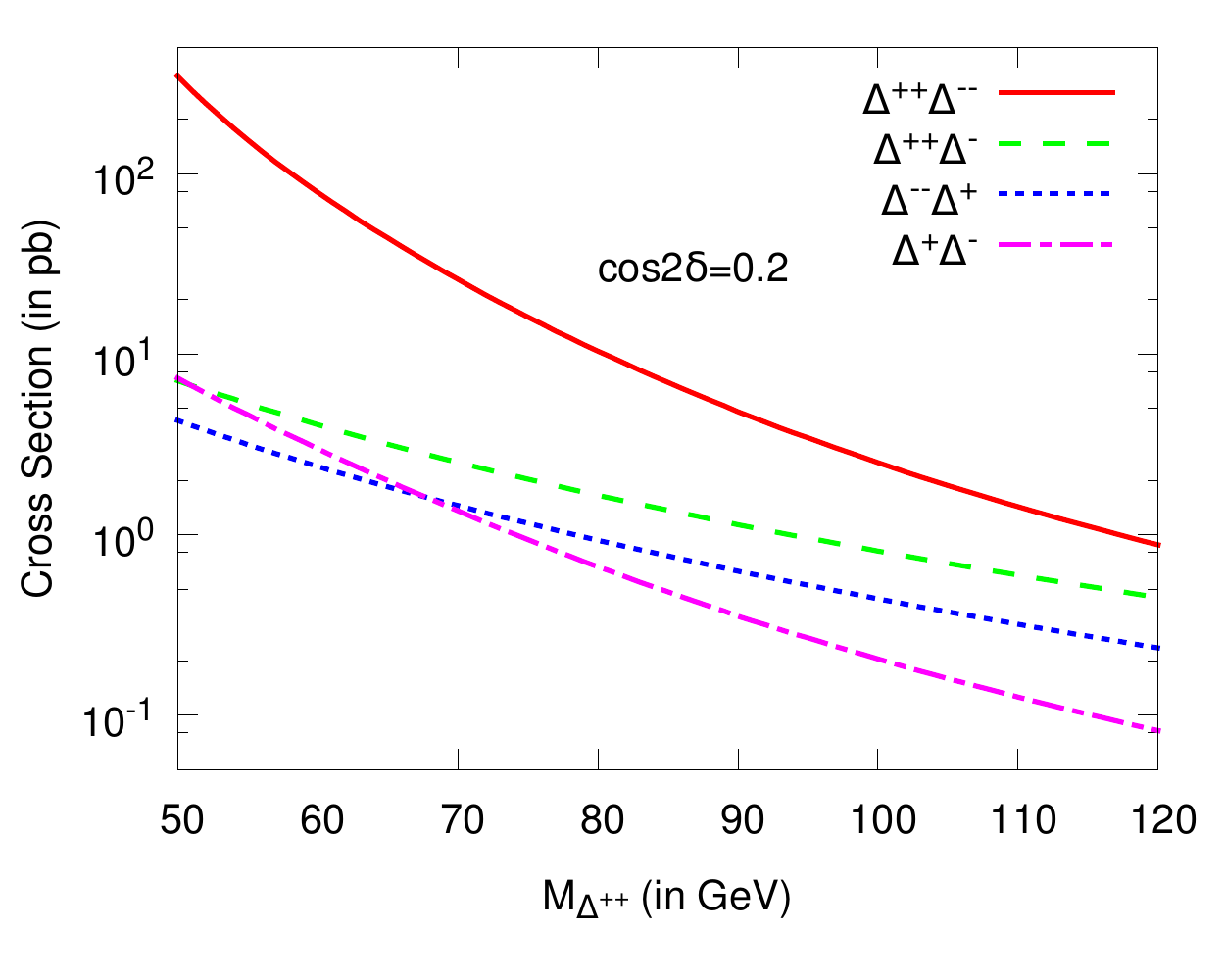}
\includegraphics[width=2.in]{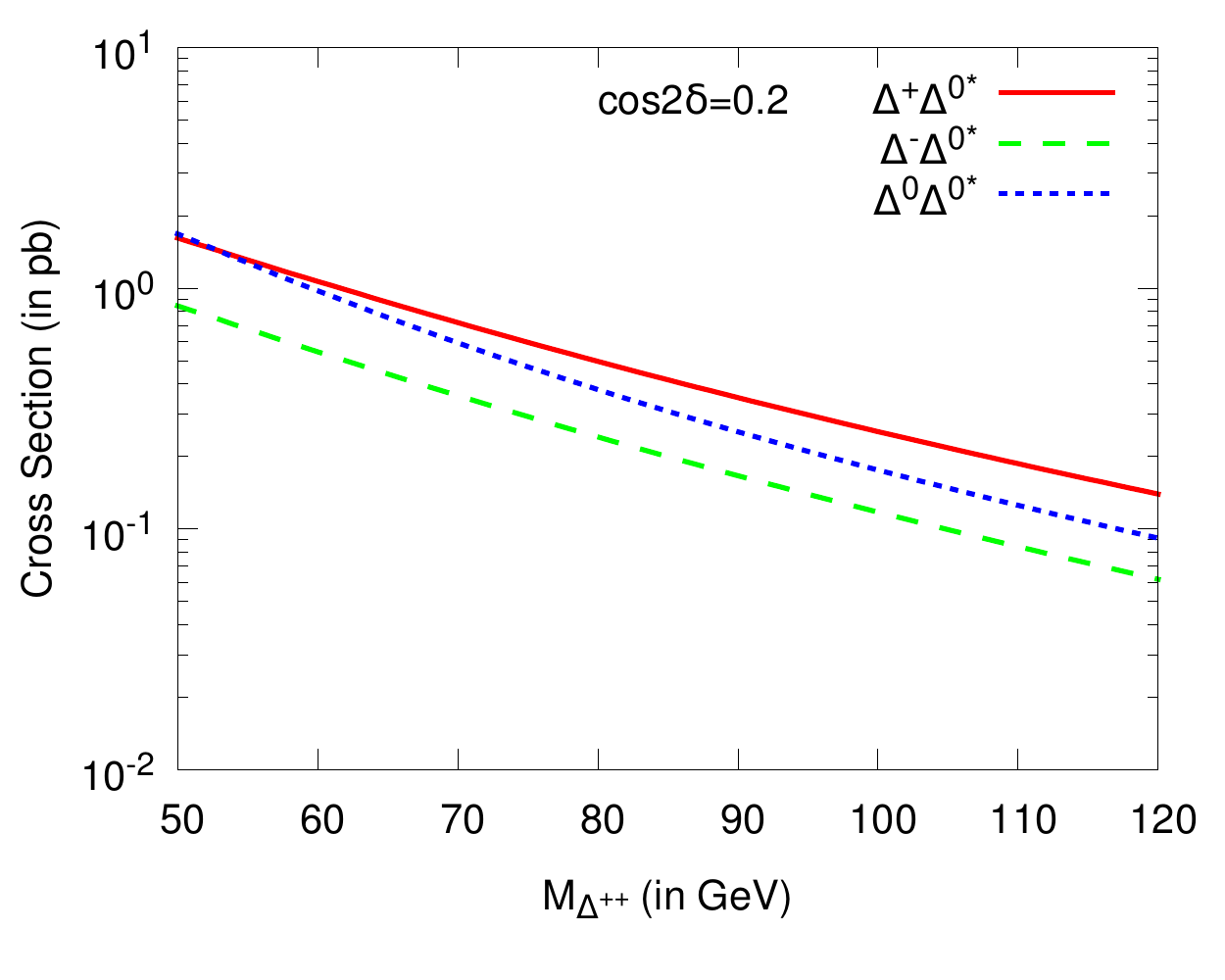}\hspace{-0.2cm}
\includegraphics[width=2.in]{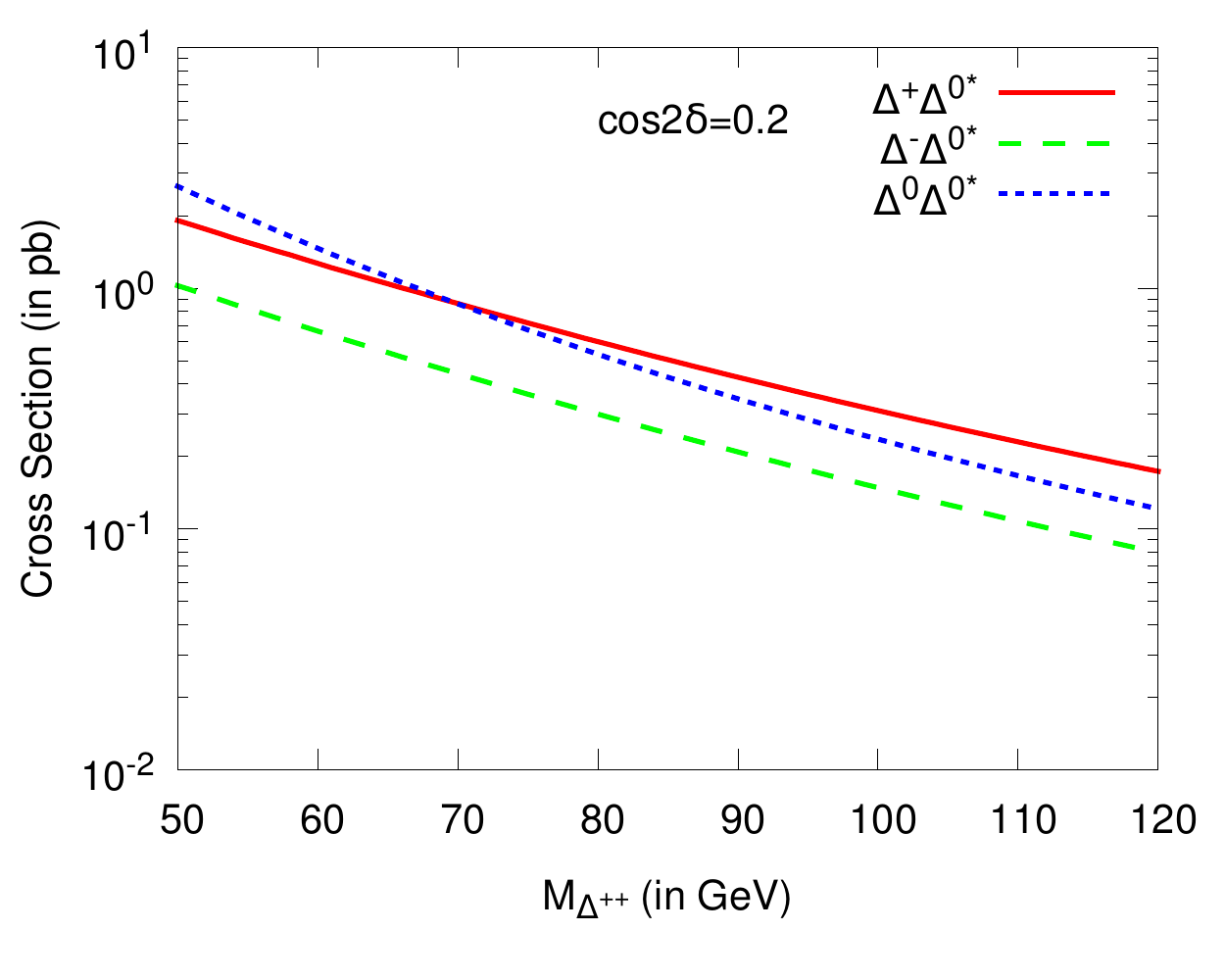}\hspace{-0.2cm}
\includegraphics[width=2.in]{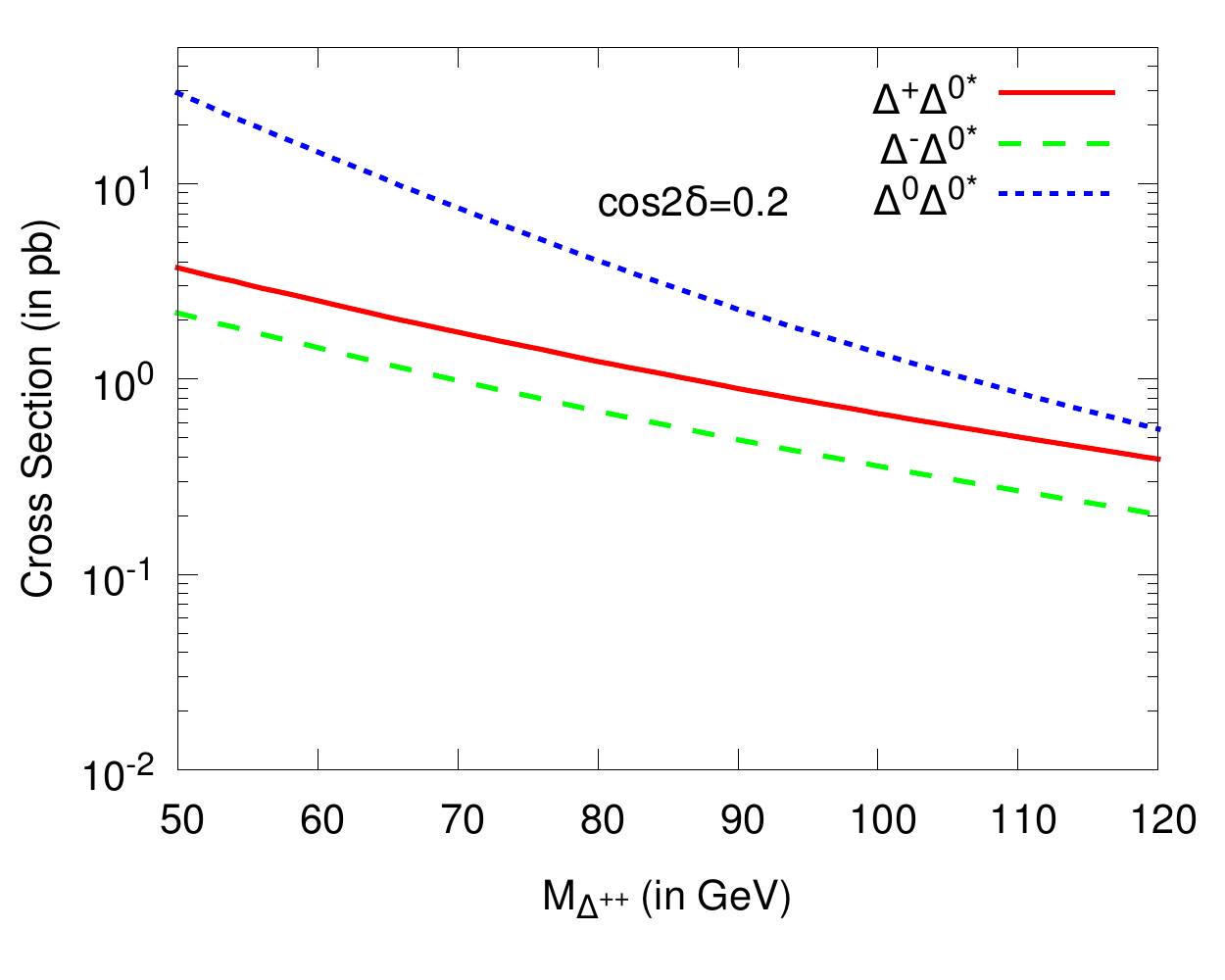}
\caption{All the pair and associated production cross-sections of triplets at 7 TeV (left), 8 TeV (middle) and 14 TeV (right) for $c_{2\delta}=$ 0.2.}
\label{crx-ch} \label{crx-2}
\end{center}
\end{figure}

In order to get projections for the LHC8 and LHC14 reach, let us first calculate
the production cross-sections of various triplet pairs.
The current search at CMS and ATLAS assumes degenerate triplet bosons which is true
for the limiting case of $c_{2\delta}=0$. However, a sizable mass splitting among the triplet components appears generically for $c_{2\delta}\neq0$ as discussed in Section 2.
In the case of $c_{2\delta}>0$, in particular, $\Delta^{++}$ becomes lighter than $\Delta^{+,0}$
and thus not only the pair production of $pp\to \Delta^{++}\Delta^{--}$ but also the gauge decays of, e.g.,  $\Delta^0\to \Delta^+ W^- \to \Delta^{++} W^- W^-$ after the associated productions of
$pp \to \Delta^{\pm\pm}\Delta^\mp, \Delta^\pm \Delta^{0(\dagger)}$ and $\Delta^0 \Delta^{0\dagger}$
can contribute to the $\Delta^{++}\Delta^{--}$ final state.
To see how sizable are the gauge decay rates of the heavier triplet boson components, we show  $\Gamma(\Delta^0\to \Delta^+ W^-)$ for various values of $c_{2\delta}$ in Fig.~\ref{gamh0}.
Comparing it with
Fig.~\ref{hppWW}, one finds that the gauge decay  is many orders of magnitude higher than
$\Gamma_{ll}$ or $\Gamma_{WW}$ for the parameter space of our interest, and thus one can infer that $\Delta^{+,0}$ will
end up with producing $\Delta^{++}$ with almost 100\% branching ratios.
We present in
Figs.~\ref{crx-8} and \ref{crx-2}  various cross-sections of the triplet boson production as functions of the doubly charged boson mass at LHC7, 8 and 14 for the
two choices of $c_{2\delta}=0.8$ and 0.2, respectively. While the pair production of $pp \to \Delta^{++} \Delta^{--}$ is larger than the other pair and associated productions, the latter are larger for smaller $c_{2\delta}$ (and thus smaller mass gap between the triplet components) and for
larger $M_{\Delta^{++}}$.

\begin{figure}[h]
\begin{center}
\hspace{-0.8cm}
\includegraphics[width=3.in]{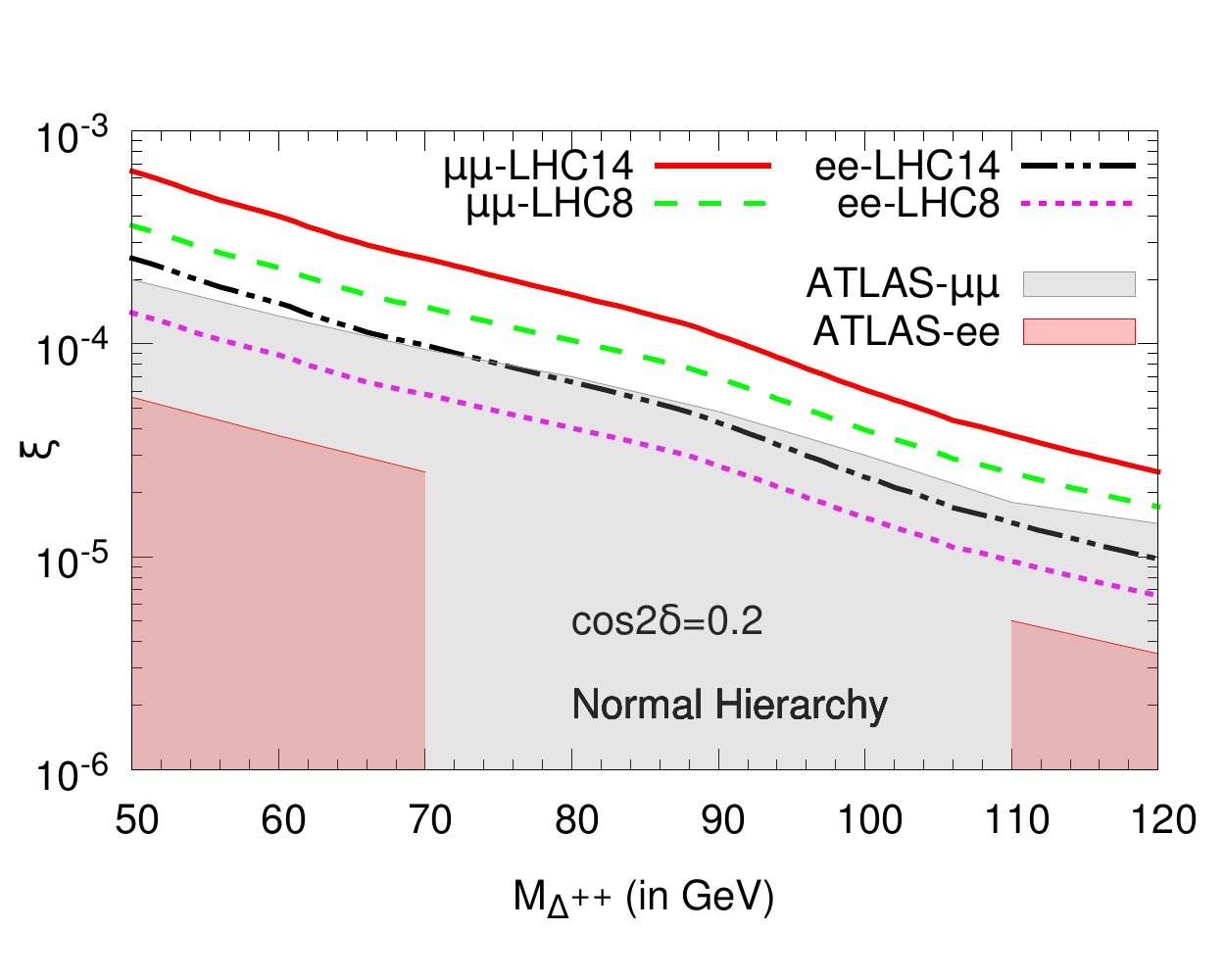} \hspace{-0.42cm}
\includegraphics[width=3.in]{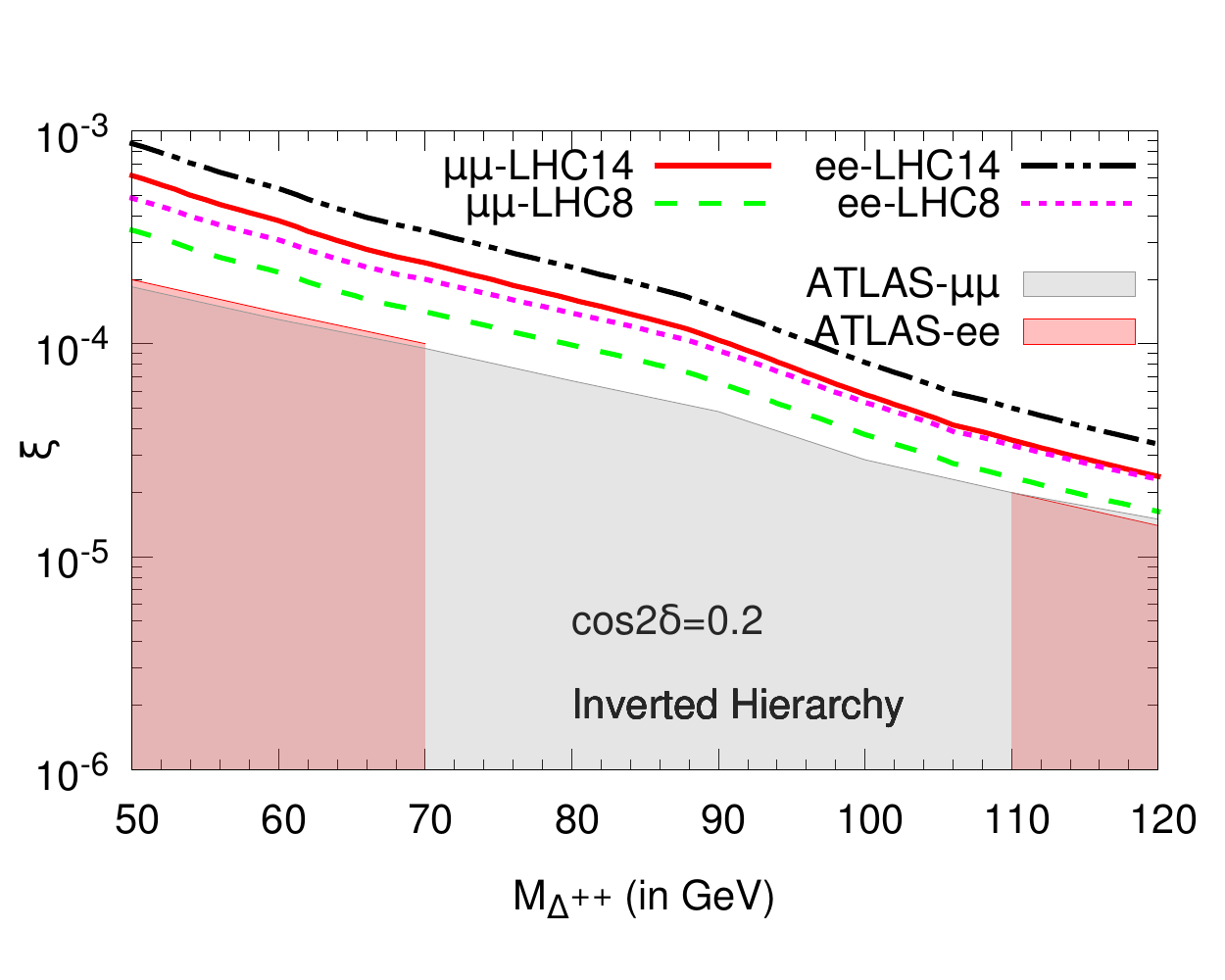} 
\caption{Cross-section 	 $\sigma(pp\to\Delta^{++}\Delta^{--}+X)\times$BF($\Delta^{\pm\pm}\to\ell^\pm\ell^\pm$) in the ($M_{\Delta^{++}}$,$\xi$)
plane for NH (left) and IH (right) at 8 TeV, 14 TeV for $c_{2\delta}=0.2$, and the ATLAS exclusion lines. The contours are for $\sigma\times$BF of 1 fb.}
\label{excl}
\end{center}
\end{figure}

Combining all possible channels for the di-lepton production,
 we plot in Fig.~\ref{excl} contour lines of  $\sigma(pp\to\Delta^{++}\Delta^{--}+X)\times\mbox{BF}(\Delta^{++}\to e^+e^+/\mu^+\mu^+)$ 
 in the ($M_{\Delta^{++}}$,$\xi$) plane at LHC8 and LHC14 together with the LHC7 exclusion lines
 for NH (left) and IH (right).
The contours denote $\sigma\times$BF of 1 fb which can be reachable readily with minimal luminosity at LHC8 and LHC14.  
One can find that the exclusion lines from LHC7 are almost same as in Fig.~\ref{brhpp} for lower mass, but a bit higher 
for heavier mass as Fig.~\ref{excl} includes more diplepton final states coming from all the associated production channels mentioned above.
We also find that
the figure for $c_{2\delta}=0.8$ changes very little as $\xi$ is insensitive to the change of $\sigma\times\mbox{BF}$ 
due to the scaling behavior of $\xi \propto 1/(\sigma\times\mbox{BF})^{1/4}$.

\medskip

Apart from the well-studied same-sign di-lepton signals, there can appear also a novel
phenomenon of same-sign tetra-leptons indicating the neutral triplet--antitriplet oscillation \cite{chun1209}. For this to occur, one needs a condition for the
oscillation parameter
\begin{equation} \label{xis}
x\equiv {\delta M \over \Gamma_{\Delta^0} }\gtrsim 1
\end{equation}
where $\delta M$ is
the mass splitting between two real degrees of freedom of the neutral triplet boson, and
$\Gamma_{\Delta^0}\simeq \Gamma(\Delta^0 \to \Delta^+ W^{-*})$.
Arising
from the lepton number violating effect, $\delta M$ is proportional to $\xi^2$ and thus can be
comparable to the decay rate of $\Gamma_{\Delta^0} \approx G_F^2 \Delta M^5/\pi^3$ which is  also
quite suppressed for a small mass gap $\Delta M \equiv M_{\Delta^0} - M_{\Delta^+}$. Precise values of $\Gamma_{\Delta^0}$ are calculated in Fig.~\ref{gamh0}
 as functions of $M_{\Delta^{++}}$ and $c_{2\delta}$.
Unlike in the non-supersymmetric type II seesaw,
$\delta M$ is a strongly model-dependent parameter in the supersymmetric version. Thus, we will parameterize its value as
\begin{equation} \label{dM}
\delta M = a \xi_1^2 M_{\Delta^0}
\end{equation}
where $a$ is an order-one parameter depending on the other model parameters
such as $\tan\beta$, the couplings $\lambda_{1,2}$ and the masses of  the heavy pseudoscalar Higgs and  triplet bosons, 
and so on. Given $M_{\Delta^{++}}, c_{2\delta}$ and $\xi=\xi_1=\xi_\Delta/c_\delta$,
one can get an estimate of $x$ from (\ref{dM}) and $\Gamma_{\Delta^0}$
in Fig.~\ref{gamh0}. For larger $c_{2\delta}$, one gets larger mass gap and thus more efficient decay of $\Delta^0 \to \Delta^+ W^{-*}$ 
suppressing the value of $x$. Once the oscillation parameter is determined, one can calculate the production cross-sections for 
the same-sign tetra-lepton final states from the following formula \cite{chun1209}:
\begin{eqnarray}
\sigma\left(4\ell^\pm + nW^{\mp^*}\right)&=&\nonumber \Bigg\{\sigma\left(pp\to
\Delta^\pm \Delta^{0(\dagger)} \right)  \left[{ x^2\over 2( 1+ x^2)}\right]
 \mbox{BF}(\Delta^{0(\dagger)}\to \Delta^\pm W^{\mp^*})\\\nonumber
&+&\sigma\left(pp\to
\Delta^0 \Delta^{0\dagger} \right)  \left[{2+x^2 \over 2(1+x^2)}
{ x^2\over 2(1+ x^2)} \right]
 \left[\mbox{BF}(\Delta^{0(\dagger)} \to \Delta^\pm W^{\mp^*})\right]^2\Bigg\} \nonumber\\
&\times& \left[\mbox{BF}(\Delta^\pm\to \Delta^{\pm\pm} W^{\mp^*})\right]^2
\left[\mbox{BF}(\Delta^{\pm\pm}\to \ell_i^\pm\ell_j^{\pm})\right]^2 .
\label{4l3W}
\end{eqnarray}

Let us now discuss
if observable same-sign tetra-lepton signals can be obtained in the parameter region
where a large enhancement of the Higgs-to-diphoton rate is obtained.
To get an idea, let us first take
an example of $M_{\Delta^{++}}=70$ GeV and $c_{2\delta}=0.8$ which gives $R_{\gamma\gamma}=1.7$.
From Fig.~\ref{gamh0}, one finds $\Gamma_{\Delta^0} \approx 10^{-4}$ GeV  and thus the oscillation for this point is
\begin{equation}
x_{(70,\, 0.8)} \approx 0.008\, a \left( 8\times 10^{-5} \over \xi_1 \right)^2
\end{equation}
inserting the value of $\xi_1$ shown in (\ref{ll/WW}). As the oscillation probability is proportional to 
a tiny number $x^2$, it is impossible to see same-sign tetra-lepton signals even at LHC14 for which the 
neutral triplet boson cross-section is just around 1 pb as shown in Fig.~\ref{crx-8}.
This feature remains true for all the parameter region enhancing the diphoton rate by more than 50 \%.
Let us remark that this conclusion can be invalidated if we relax the condition of $a\sim 1$ and
$\xi_1 \sim \xi_\Delta/c_\delta$ to allow a large deviation from this generic relation accepting
a certain fine-tuning of parameters.

A larger oscillation probability can be obtained for smaller $c_{2\delta}$ (and thus smaller $\Delta M$)
suppressing the decay rate $\Gamma_{\Delta^0}$.  To see this, let us now take  $M_{\Delta^{++}}=70$ GeV with $c_{2\delta}=0.2$ which gives $\Gamma_{\Delta^0} \approx 2\times
10^{-7}$ GeV and
\begin{equation}
x_{(70,\, 0.2)} \approx 2.8\, a \left( 8\times 10^{-5} \over \xi_1 \right)^2 .
\end{equation}
Considering the pair and associated production cross-section in Fig.~(\ref{crx-2}) and
the leptonic branching fraction $\sim 1$ \% as in (\ref{ll/WW}), one gets from (\ref{4l3W})
the cross-section $\sim 0.3$ fb for the same-sign tetra-lepton production at LHC14.
Thus, same-sign tetra-lepton signals can be observable for the integrated luminosity larger than 10 fb$^{-1}$ at LHC14.
 Note that, this sample parameter point with $M_{\Delta^{++}}=70$ GeV can be easily accessible
  through the same-sign di-lepton search as its production cross-section
dominated by the pair production of $\Delta^{++} \Delta^{--}$ is about 20 pb which is
20 times larger than in LHC7 or LHC8.


\section{Conclusion}

The supersymmetric type II seesaw may exhibit a `fine-tuned' possibility allowing only one of
the triplet bosons and the standard Higgs boson as light degrees of freedom
below the TeV scale around which supersymmetry is supposed to be broken.
In this limit, the particle content is the same as
in the non-supersymmetric model, but there appear more parameters, depending on
the masses spectrum and couplings of heavy particles, which complicates the collider phenomenology
of the light triplet bosons compared to the non-supersymmetric type II seesaw.
However, the mass splitting among the triplet components and
the triplet-Higgs couplings are determined simply by the $D$-term potential
neglecting contributions from triplet vacuum expectation values which are assumed to be small.

The doubly charged boson which can be the lightest among three triplet components may contribute
significantly to the Higgs decay to diphoton.  Coming from the $D$ term,
the trilinear coupling of the light doubly (and singly) charged boson to the Higgs boson
is smaller than in the non-supersymmetric model, and its mass needs to be smaller
than 80 GeV if the Higgs-to-diphoton rate is to deviate from the Standard Model
prediction by larger than 50 \% [Fig.~\ref{Rgg}].
Such a light doubly charged boson can evade the current LHC search
if its leptonic branching ratio is small enough, but will be probed with accumulating data.
It gives another interesting possibility of another unconventional Higgs decay 
to a pair of light doubly charged bosons which would be
related to the non-standard Higgs-to-diphoton rate.

Including all the pair and associated productions of the triplet pairs with the mass hierarchy
of $M_{\Delta^0} > M_{\Delta^+} > M_{\Delta^{++}}$, the excluded region in the parameter space
$(M_{\Delta^{++}}, \xi)$ is obtained from the ATLAS analysis of LHC7 data on the same-sign di-lepton channels, and LHC8 and LHC14 projections are derived in Fig.~\ref{excl}.
In addition to the usual di-lepton signals, the type II seesaw model can be tested also
by observing same-sign tetra-leptons arising from the triplet-antitriplet oscillation.
However, the cross-sections for such a signature in the parameter region
allowing a sizable deviation of the diphoton rate turn out to be too small
to be probed even at the LHC14.  Thus, observation of such a signal would exclude the triplet bosons from contributing to a possibly large deviation of the Higgs-to-diphoton rate.

\appendix
\section{}

The scalar potential of the supersymmetric type II seesaw model consists of the contributions
from the supersymmetric $F$ and $D$ terms as well as the soft supersymmetry breaking terms which are given by
\begin{eqnarray} \label{VFDs}
 V_F &=& | {\lambda_1\over2} H_1^0 H_1^0 + M \bar{\Delta}^0|^2
 + |{\lambda_2\over2} H_2^0 H_2^0 + M \Delta^0|^2 \\
&& + | \lambda_1 H_1^0 \Delta^0 -{\lambda_1\over \sqrt{2}} H_1^- \Delta^+ + \mu H_2^0|^2
   + | \lambda_2 H_2^0 {\bar\Delta}^0 +{\lambda_2 \over \sqrt{2}} H_2^+ \bar\Delta^- + \mu H_1^0|^2 \nonumber\\
&& + |{\lambda_1\over\sqrt{2}} H_1^0 \Delta^+ + H_1^- \Delta^{++} + \mu H_2^+|^2
   + |{\lambda_2\over\sqrt{2}} H_2^0 \bar\Delta^- - H_2^+ \bar\Delta^{--} - \mu H_1^-|^2 \nonumber\\
&& + |{\lambda_1\over\sqrt{2}}H_1^0 H_1^- - M \bar\Delta^-|^2
   + |{\lambda_2\over\sqrt{2}}H_2^0 H_2^+ + M \Delta^+|^2 \nonumber\\
&& + | {\lambda_1\over\sqrt{2}} H_1^- H_1^- - M \bar\Delta^{--}|^2
   + | {\lambda_2\over\sqrt{2}} H_2^+ H_2^+ - M \Delta^{++}|^2 \,,
 \nonumber\\
 V_D &=& {g^2\over 8} \left[ |H_1^0|^2-|H_2^0|^2 + 2|\Delta^{++}|^2 - 2
 |\Delta^{0}|^2 - 2|\bar{\Delta}^{--}|^2  + 2|\bar{\Delta}^{0}|^2\right]^2
 \\
  &+&
 {g^{'2}\over 8} \left[ |H_1^0|^2 - |H_2^0|^2 - 2|\Delta^{++}|^2 - 2|\Delta^{+}|^2
  -2 |\Delta^{0}|^2 +  2|\bar{\Delta}^{--}|^2  + 2|\bar{\Delta}^{-}|^2 + 2|\bar{\Delta}^{0}|^2\right]^2 \,, \nonumber\\
 V_{soft} &=& + {1\over2} f_1 A_1
 \left[ H_1^0 H_1^0 \Delta^{0} -\sqrt{2} H_1^0 H_1^- \Delta^+
 - H_1^- H_1^- \Delta^{++} + h.c. \right] \\
&&  + {1\over2}f_2 A_2
 \left[H_2^0 H_2^0 \bar\Delta^{0} + \sqrt{2} H_2^0 H_2^+ \bar\Delta^-
 - H_2^+ H_2^+ \bar\Delta^{--} + h.c. \right]
 \nonumber\\
&& + B_\mu \left[ H_1^0 H_2^0 - H_1^- H_2^+ + h.c.\right]
+ B_M \left[ \Delta^a \bar\Delta^{\bar a}  + h.c. \right]
 + m_{\Delta}^2 |\Delta^a|^2  +  m_{\bar\Delta}^2  |\bar{\Delta}^{\bar a}|^2 \,.
 \nonumber
\end{eqnarray}
Note that only the $T_3$ term of $SU(2)_L$ and the $Y$ term of $U(1)_Y$ are shown in $V_D$, and the index $a$ runs for the triplet components ($a=++, +,0$ and $\bar a = --, -, 0$).  It is assumed that there is no $CP$ phase in the couplings.\\

\textit{Acknowledgements:} EJC was supported by the National Research Foundation of Korea (NRF) grant funded by the Korea government (MEST) (No.~20120001177).  We thank Zhaofeng Kang for communications on the Higgs decay and on the new results appeared in \cite{kang13}.

\end{document}